\documentclass[sigconf,nonacm]{acmart}

\AtBeginDocument{%
  }

\setcopyright{acmlicensed}
\copyrightyear{2018}
\acmYear{2018}
\acmDOI{XXXXXXX.XXXXXXX}

\acmConference[Conference acronym 'XX]{Make sure to enter the correct
  conference title from your rights confirmation email}{June 03--05,
  2018}{Woodstock, NY}

\acmISBN{978-1-4503-XXXX-X/2018/06}

\usepackage{graphicx} 
\usepackage{xspace}
\usepackage{algorithm}
\usepackage{tablefootnote}
\usepackage{algpseudocode}
\usepackage{adjustbox}
\usepackage{threeparttable,booktabs}
\usepackage{siunitx}
\usepackage{mhchem}
\usepackage{pifont}
\usepackage{multirow}
\usepackage{graphicx}
\usepackage{ulem}
\usepackage{subcaption}
\usepackage{xcolor}
\usepackage{tabularx}  
\usepackage{booktabs}
\usepackage{multicol}
\usepackage{comment}
\usepackage{url}
\usepackage{mdframed}

\newcommand{\dong}{\textcolor{blue}}
\newcommand{\name}{\textsc{DOLMA}\xspace}

\newcommand{\summary}[1]{
\noindent
\begin{mdframed}[linecolor=black, linewidth=0.5pt, backgroundcolor=lightgray!20,
skipabove=-5pt, 
skipbelow=-1pt  
  ]
\textbf{Key Takeaway: } #1
\end{mdframed}
}
\usepackage{xifthen}

\definecolor{custompurple}{rgb}{0.435, 0.0, 1.0}

\begin{document}
\pagestyle{plain}  
\thispagestyle{plain}  

\title{\name: {A Data Object Level Memory
Disaggregation Framework for HPC Applications}}

\author{%
Haoyu Zheng$^{1}$,
Shouwei Gao$^{1}$,
Jie Ren$^{2}$,
Wenqian Dong$^{1}$ \\
{\small $^{1}$Oregon State University, Corvallis, OR, USA \quad
$^{2}$College of William \& Mary, Williamsburg, Virginia, USA \quad}
}

\begin{abstract}
  Memory disaggregation is promising to scale memory capacity and improves utilization in HPC systems. However, the performance overhead of accessing remote memory poses a significant challenge, particularly for compute-intensive HPC applications where execution times are highly sensitive to data locality. In this work, we present \name, a \underline{D}ata \underline{O}bject \underline{L}evel \underline{M}emory dis\underline{A}ggregation framework designed for HPC applications. 
  \name intelligently identifies and offloads data objects to remote memory, while providing quantitative analysis to decide a suitable local memory size. Furthermore, \name leverages the predictable memory access patterns typical in HPC applications and enables remote memory prefetch via a dual-buffer design. 
  By carefully balancing local and remote memory usage and maintaining multi-thread concurrency, \name provides a flexible and efficient solution for leveraging disaggregated memory in HPC domains while minimally compromising application performance. 
  Evaluating with eight HPC workloads and computational kernels, \name limits performance degradation to less than 16\% while reducing local memory usage by up to 63\%, on average.

\end{abstract}

\keywords{Memory disaggregation, Remote memory}

\maketitle

\section{Introduction} 


  
Memory underutilization and imbalance are significant challenges in High-Performance Computing (HPC) systems, leading to increased costs and inefficient resource management~\cite{sc23_quantitative}. Studies have shown that in many HPC environments, a large portion of memory resources remains underutilized. For instance, NERSC's Cori supercomputer observed that only 15\% of scientific workloads utilize over 75\% of available memory per node~\cite{ivy20_hpc_mem_underutilization}. At Lawrence Livermore National Laboratory, 90\% of jobs use less than 15\% of node memory capacity~\cite{peng2020memory}. Furthermore, up to 83\% of memory can be underutilized on tightly-coupled resources that are over-provisioned for workloads with the highest demands~\cite{ding2022methodology}. 
These studies reveal a critical mismatch between fixed per-node memory allocation and the diverse memory requirements of HPC workloads. 

Memory disaggregation offers a promising solution to these challenges by decoupling directly accessible memory from monolithic servers and enabling compute nodes to access memory from remote memory nodes. 
This approach allows for improved memory capacity through disaggregation, cost savings, separate scaling of processing and remote memory resources, and better overall memory resource utilization. Recent advancements in hardware technologies, such as Remote Direct Memory Access (RDMA)~\cite{rdma}, new interconnects~\cite{infiniband_2024} and switches\cite{poon2009cascaded}, have made memory disaggregation increasingly feasible in HPC environments. With InfiniBand networks now capable of achieving speeds up to 800Gb/s and sub-600 nanosecond latency~\cite{infiniband_2024}, the performance gap between local and remote memory access has narrowed significantly, making memory disaggregation a realistic and beneficial approach for HPC clusters.

Existing research on disaggregated memory systems often relies on a Unified Address Space (UAS) approach with memory page-level management~\cite{osdi18_legoos,  sigmod22_teleport, nsdi17_infiniswap, bigdata18_fastswap, atc20_clover, ATC20_leap}. While UAS provides ease of use for application developers, it introduces substantial overhead in address translation and causes significant data access amplification, particularly for data-intensive HPC applications characterized by large and complex memory access patterns. The fine-grained nature of page-level management often results in excessive remote memory accesses, degrading overall performance.

Some studies explore data object-level memory disaggregation~\cite{osdi22_carbink, fast22_Hydra, Lee2019MitigatingTP, Keeton2021MODCRF, OSDI20_AIFM} in transaction-based applications. These solutions aim to reduce the overhead associated with remote memory access by operating at a coarser granularity and considering the semantic boundaries of data structures. However, existing data object-level approaches are primarily tailored for commercial databases, which have different access patterns and requirements compared to HPC workloads. Database applications typically involve well-defined, structured data queries and transactions, whereas HPC applications often deal with complex, irregular, and data-dependent access patterns. Consequently, there is a critical gap in research on memory management solutions specifically designed for the unique challenges of complex, data-intensive HPC applications.

Current memory disaggregation solutions, from page-level systems to object-level designs for transactional workloads, have shown promising results. However, adapting memory disaggregation for HPC applications poses unique challenges due to their computational intensive nature and strict requirements for memory bandwidth and latency.
First, HPC data objects vary significantly in size and often don't align with memory page boundaries or local buffer sizes. This misalignment complicates decisions about when to fetch and evict data objects, directly impacting remote memory access performance. Unlike transactional workloads with more predictable object sizes, HPC applications must efficiently manage these diverse data objects across the memory hierarchy. 
Second, the irregular access patterns in HPC applications make synchronization particularly challenging in disaggregated memory environments. Ensuring application correctness while leveraging asynchronous RDMA operations to maximize performance requires complex coordination mechanisms. 
Third, these challenges compound in multi-threaded environments, where concurrent access to shared data objects requires careful coordination across threads and nodes to maintain performance at scale.



In this work, we present \name, a data object-level memory disaggregation framework designed for HPC applications. By implementing \name on top of the Infiniband network architecture, we enable significant local memory savings without incurring large performance degradation. 
First, to address the challenge of data object and local memory buffer mismatch, \name leverages the unique characteristics of HPC applications to intelligently partition and place data objects. Our framework places large, long-lived data objects in dedicated memory nodes while keeping frequently accessed, small, and short-lived data objects in local compute node memory. This strategic placement minimizes data movement and reduces remote memory access overhead. 
Second, to tackle synchronization challenges, \name employs a dual buffer design and carefully relaxes synchronization for remote write operations. By leveraging the iterative nature of HPC computations, our framework hides data movement overhead by overlapping it with computation, minimizing the impact on application performance. 
Third, to address multi-threading challenges, \name implements per-thread local buffers based on OpenMP's parallel programming model, enabling efficient coordination of shared data object access across threads while maintaining application scalability.

We summarize the main contributions of \name as follows:
\begin{itemize}
\item A systematic analysis of HPC data object access patterns and remote memory overhead, revealing insights that guide the design of efficient memory disaggregation systems.
\item Design and implementation of \name, a memory disaggregation framework for efficient data object level remote access in HPC applications.
\item Real-system evaluation of \name on RDMA-enabled disaggregated memory using eight representative HPC applications, demonstrating 63\% reduction in local memory requirements while maintaining performance degradation within 16\% of monolithic system.
\end{itemize}

\section{Background}

\subsection{Disaggregated Memory  and HPC Workloads}

\begin{figure}[H]
    \centering
    \includegraphics[width=0.4\textwidth]{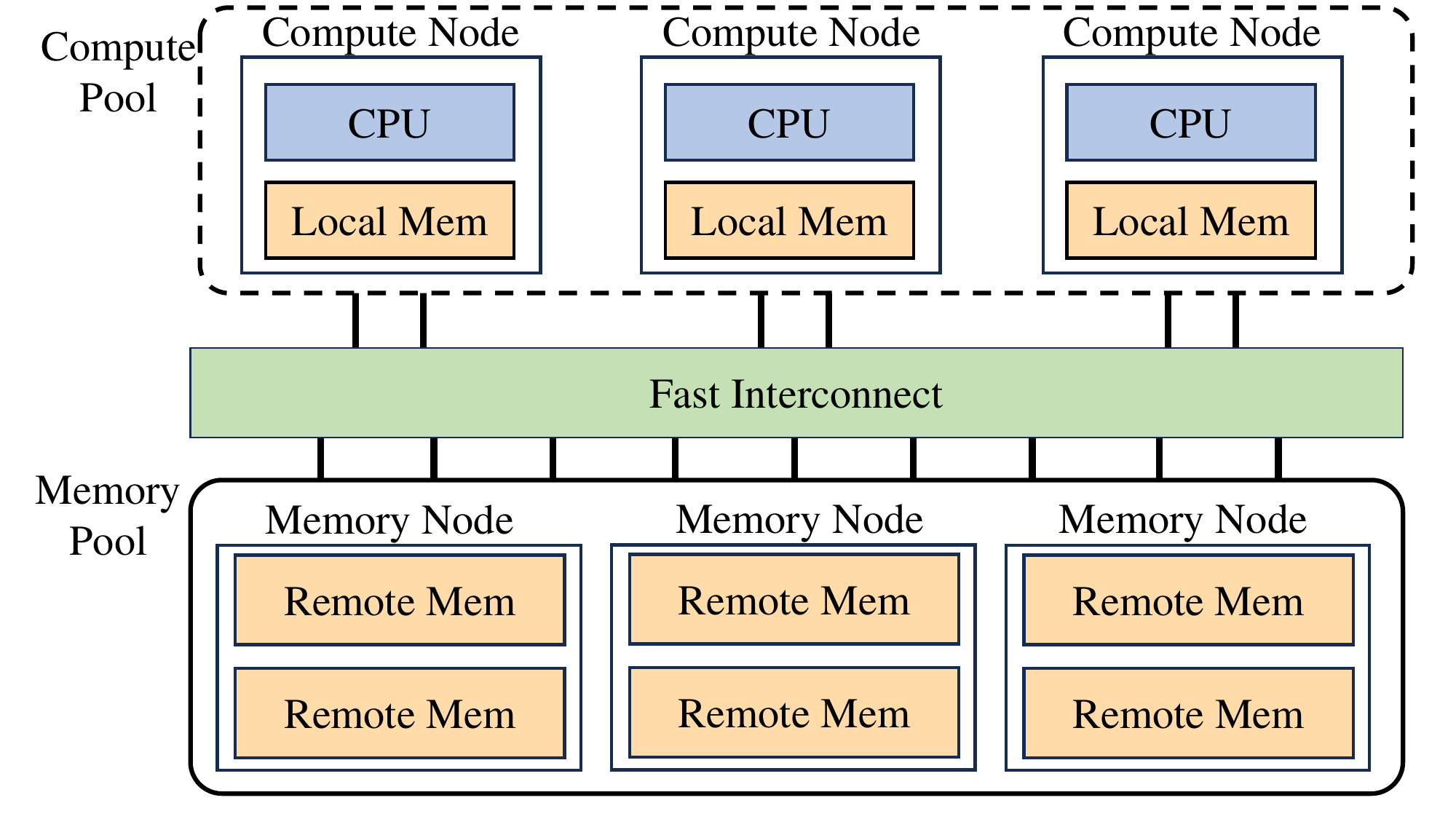}
    \vspace{-5pt}
    \caption{Disaggregated memory architecture}
        \vspace{-5pt}
    \label{fig:remote_mem}
\end{figure}

Disaggregated memory architecture decouples memory from compute nodes, enabling memory to reside in separate memory nodes connected via high-speed interconnects. This separation allows memory and compute to scale independently and supports more flexible system architectures. As shown in Figure~\ref{fig:remote_mem}, each compute node includes local memory but can also access remote memory through interconnects such as InfiniBand or CXL.

Prior work~\cite{lim2009disaggregated, nitu2018welcome, novakovic2014scale, nsdi17_infiniswap, osdi18_legoos} has explored transparent access to remote memory by extending the operating system with page-level swapping mechanisms. While this approach simplifies application integration, it incurs significant overhead due to page fault handling, kernel involvement, and coarse-grained memory tracking, which lead to high access latency and CPU contention.

Meanwhile, many modern cloud applications, such as data analytics frameworks~\cite{fast21_tectonic}, web services~\cite{ATC20_leap, firecracker}, and key-value stores~\cite{sosp23_ditto, OSDI20_AIFM}, benefit from object-level remote memory access. These workloads typically involve limited computation, have small data footprint per access, tolerate higher memory latency through batching or caching, and often access memory in coarse or application-managed granularities. Cloud applications can effectively hide remote memory latency and exploit disaggregation to improve resource utilization and elasticity.
  
Compared to cloud and data analytics workloads, HPC applications present distinct and more demanding challenges for disaggregated memory systems. HPC workloads are typically \textit{compute-intensive, highly parallel}, and exhibit tight coupling between computation and memory access. Their performance is often tightly bound to memory latency and bandwidth, to the extent that the additional delays and variability introduced by remote memory access can cause severe slowdowns, making both page-level and naive data object-level remote memory access unacceptable for many HPC applications.

At the same time, HPC workloads often exhibit \textit{predictable and structured memory access patterns}, such as regular strides, spatial locality, and well-defined access sequences across loop iterations or time steps. These characteristics differentiate them from the more irregular and loosely coupled memory behavior observed in many cloud workloads. This predictability creates an opportunity for fine-grained remote memory management through techniques such as prefetching, fine-grained object tracking, and workload-aware memory placement, which can make disaggregated memory practical for HPC applications.

\begin{figure}[htbp]
    \centering
    \includegraphics[width=0.45\textwidth]{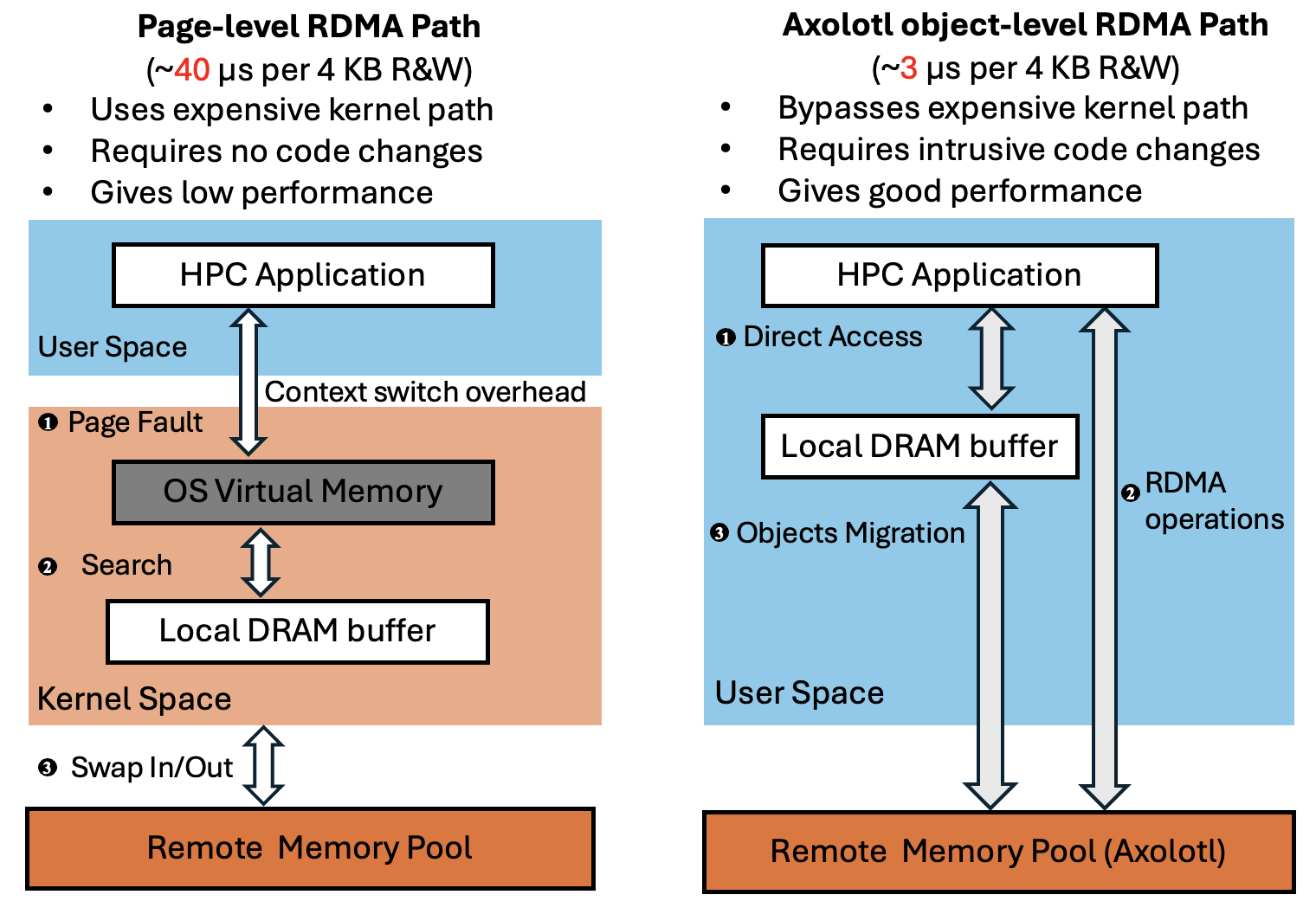}
    \vspace{-5pt} 
    \caption{Architectures of different disaggregrated memory systems. }
    \label{fig:RDMA}
    \vspace{-10pt} 
\end{figure}

\subsection{One-side RDMA Communication}
RDMA offers versatile connection options for either one-sided or two-sided communication, including \textit{reliable connections},  unreliable connections, and unreliable datagrams, to either ensure data integrity or prioritize low latency.  

\begin{figure}[H]
    \centering
    \includegraphics[width=0.45\textwidth]{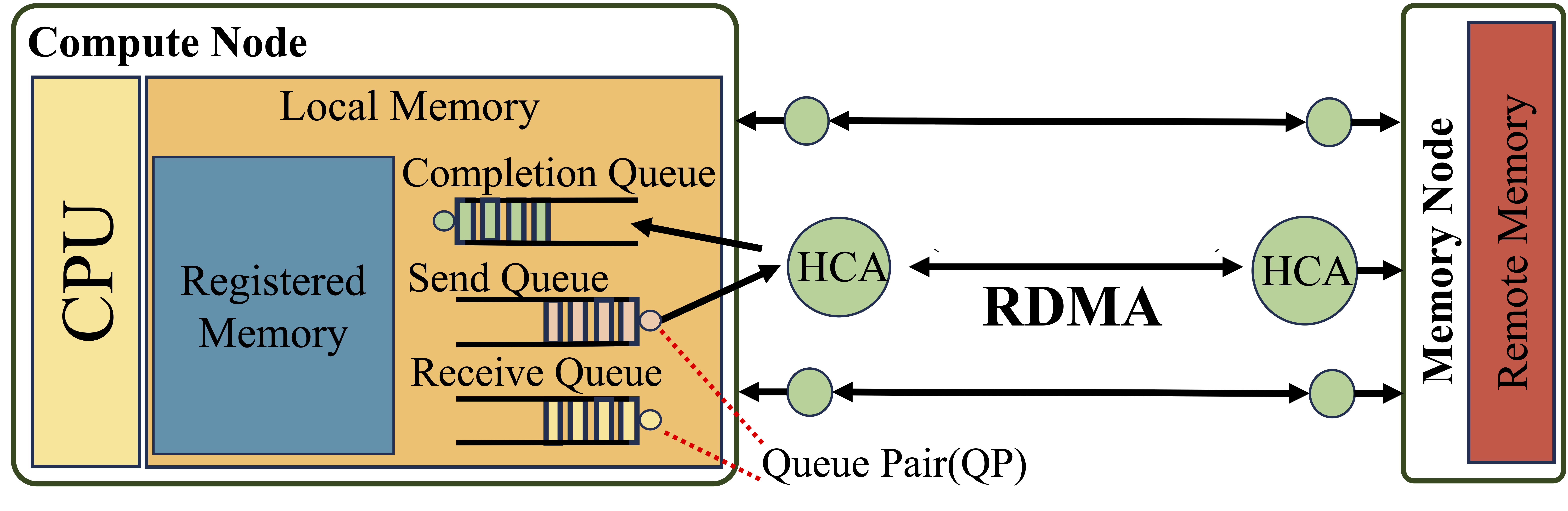}
    \vspace{-5pt}
    \caption{Example of one-side RDMA access. }
    \label{fig:RDMA}
\end{figure}

The reliable connection type establishes a \textit{queue pair} (QP) between the two communicating nodes (i.e., a local node and a remote node). The QPs are consisting of a \textit{send queue} and a \textit{receive queue}, and post operations to these queues using the verbs API.
Upon the completion of a verb, the local node’s RNICs optionally signals completion by DMAing a \textit{completion queue entry} (CQE) to a \textit{completion queue} associated with the QPs.
Endpoints of a single reliable connection QPs can only communicate with each other, but not with any other QP in the same or any other target adapter. Each QP endpoint has a \textit{queue pair number} (QPN) assigned by the RNIC which uniquely identifies the QP within the RNIC.

The memory protection mechanisms provide protection from unauthorized access to the local memory by network controllers. The local memory can also be protected against prohibited memory accesses. Three mechanisms exist to enforce memory access restrictions: Memory Regions, Protection Domains, and Memory Windows. To get access to host memory, the RNIC must first allocate a corresponding \textit{memory region}. This process involves copying page table entries of the corresponding memory to the memory management unit of the RNIC.

\textbf{An example.} Figure~\ref{fig:RDMA} shows how a compute node accesses the memory node through one-sided RDMA communication.  
%
\textit{(1) Connection Setup.} The compute node first establishes a reliable connection to the memory node by configuring a QP on the Host Channel Adapter (HCA). This QP consists of a send queue for issuing requests and a receive queue for receiving incoming data. The memory node prepares its memory region by registering it with the RNIC and granting access permissions.
\textit{(2) Read and Write Processing.} The compute node submits memory read and write requests to the memory node by posting work requests to the send queue of the QP. The RNIC's drivers will push these requests through RDMA and map the memory region from the remote node to its local memory, without involving the memory node's CPU.
\textit{(3) Completion.} Upon completing the data transfer, the memory node's RNIC sends an acknowledgment or completion notification back to the compute node. The RNIC at the compute node places a CQE into the completion queue to signal the completion of a work request.

One-sided communication involves only the compute unit 
on the local node to initiate data transfer, without requiring the participation of the remote CPU. Although the local NIC enqueues the RDMA operation in its work queue, the remote node remains passive, with no involvement of its request queue.

\begin{figure*}[!ht]
    \centering
        \begin{subfigure}[t]{0.47\textwidth}
        \captionsetup{skip=-1pt} 
        \includegraphics[width=\textwidth]{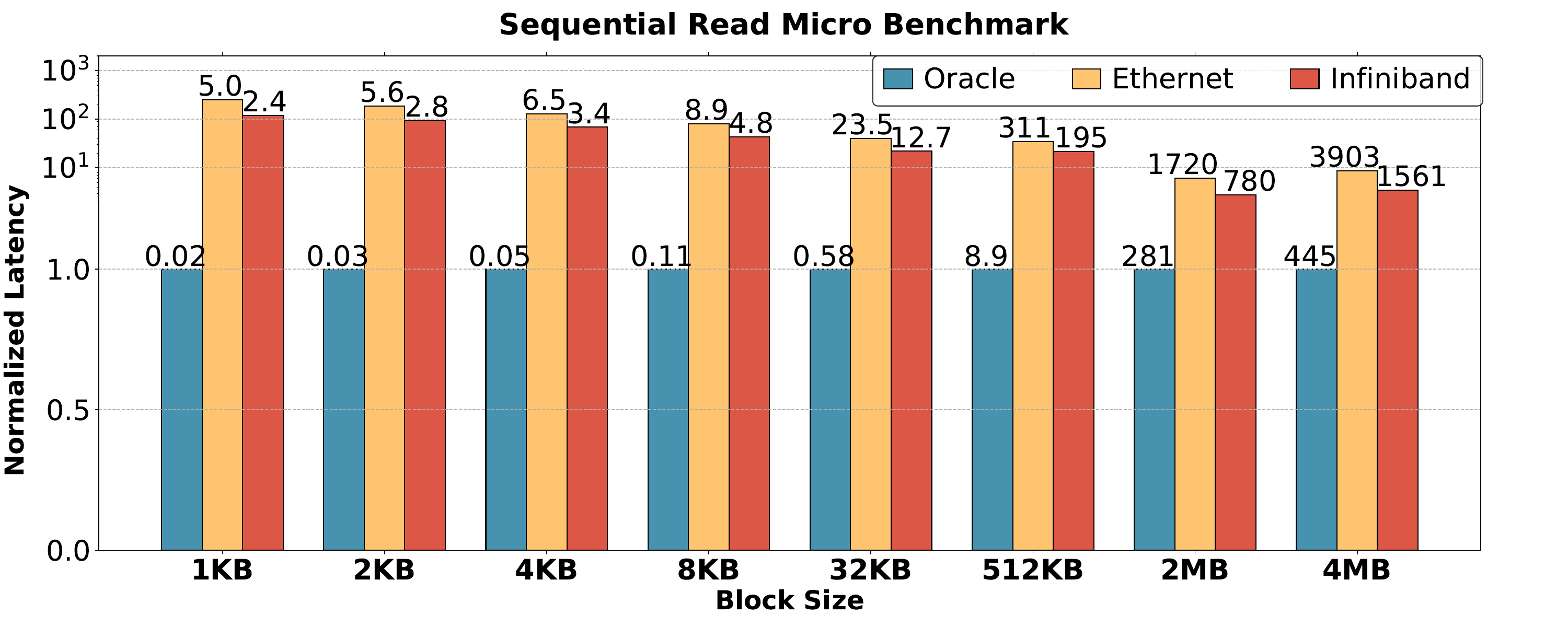}
        \caption{Sequential read latency on local and remote memory}
    \end{subfigure}
    \hspace{0.01\textwidth}
    \begin{subfigure}[t]{0.47\textwidth}
        \captionsetup{skip=-1pt}
        \includegraphics[width=\textwidth]{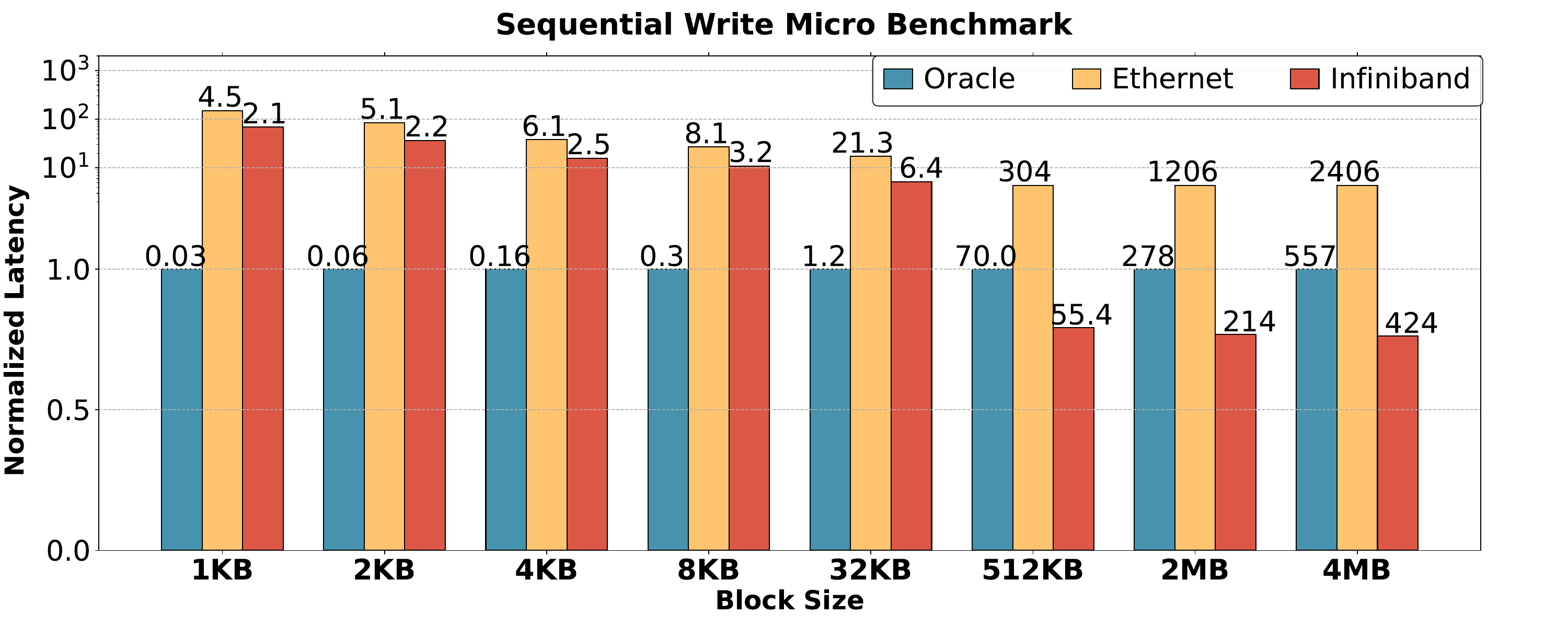}
        \caption{Sequential write latency on local and remote memory}
    \end{subfigure}
    \medskip
    \begin{subfigure}[b]{0.47\textwidth}
        \captionsetup{skip=-1pt}
        \includegraphics[width=\textwidth]{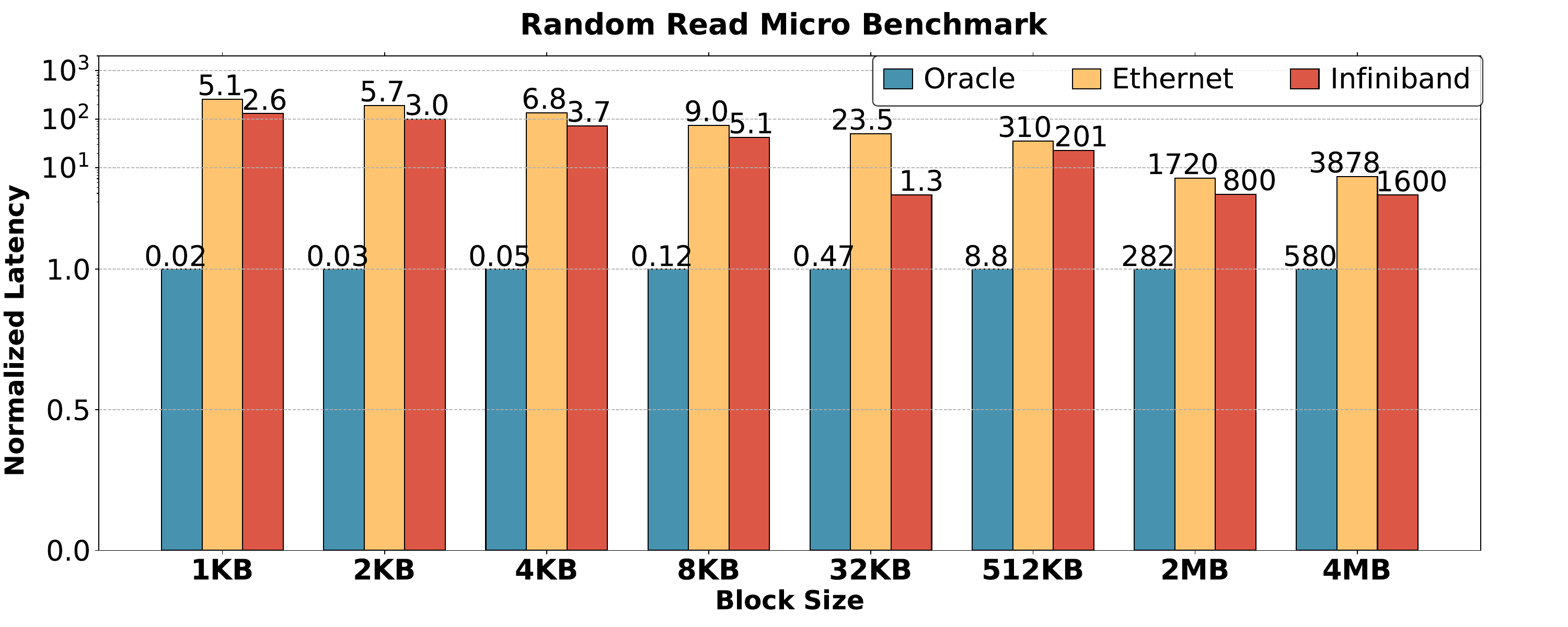}
        \caption{Random read latency on local and remote memory}
    \end{subfigure}
    \hspace{0.01\textwidth}
    \begin{subfigure}[b]{0.47\textwidth}
        \captionsetup{skip=-1pt}
        \includegraphics[width=\textwidth]{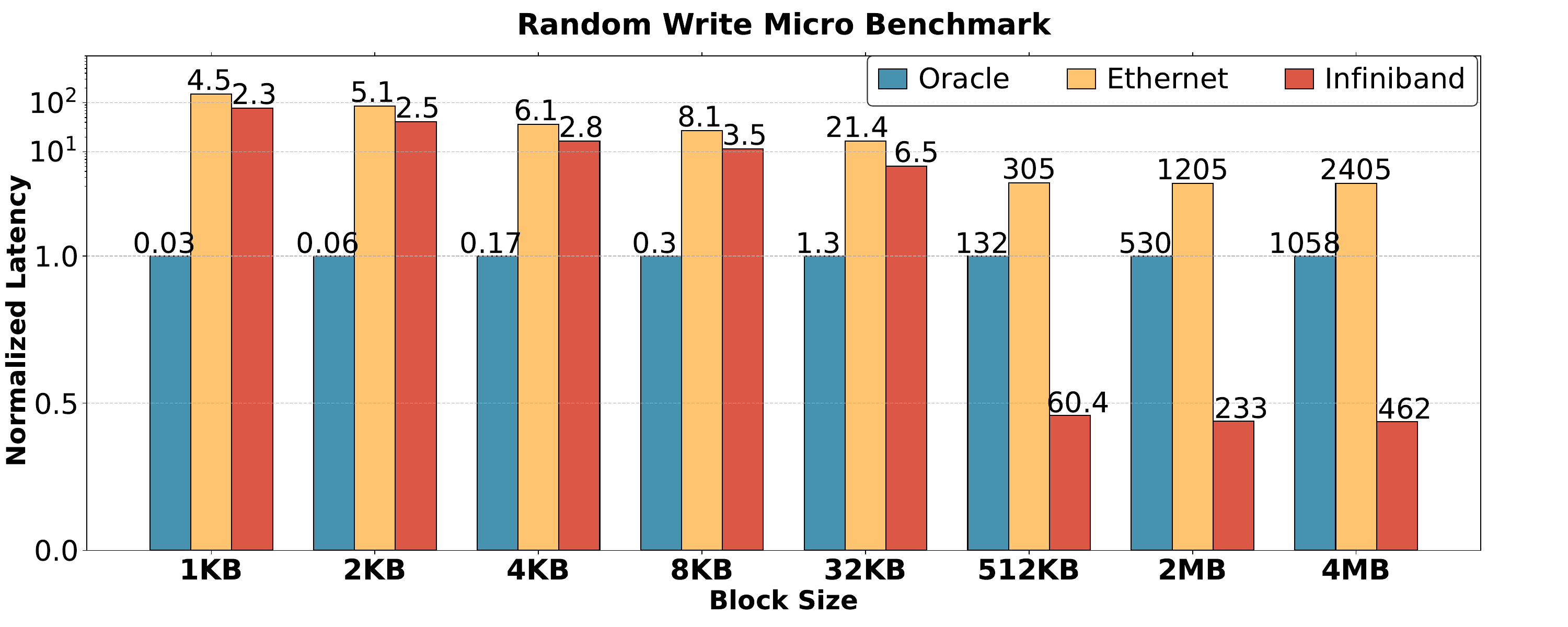}
        \caption{Random write latency on local and remote memory}
    \end{subfigure}
    \vspace{-10pt}
    \caption{Normalized performance of remote memory accesses with various data object sizes. The performance is normalized to local memory accesses (oracle), and the remote access latency via Infiniband (in microseconds) is labeled above the red bar.}
    \label{fig:motivation1}
\end{figure*}

\section{Preliminary Study and Motivation}

\label{sec:motivation}

%
We conduct our preliminary study on an RDMA-based disaggregated system. To simplify the scenario, we set up only two nodes: the compute node, considered as a local node, primarily performs computation-intensive workloads that require frequent data accesses; the memory node is considered as the remote node, connected via InfiniBand (with RDMA) or Ethernet.
Both nodes are equipped with two 24-core Intel(R) Xeon(R) CPUs and 187 GB of DRAM.
We set up three settings to evaluate local and remote memory access:

\texttt{\textbf{(1) Oracle:}} A setting where the compute node can access only its local memory (with NUMA), and no disaggregated memory is available; 

\texttt{\textbf{(2) Ethernet:}} A setting where the compute node accesses remote memory using RDMA over Ethernet (25 Gbps);

\texttt{\textbf{(3) InfiniBand:}} The one where the compute node can access remote memory using RDMA over InfiniBand (100 Gbps).






\subsection{Characteristics of Remote Memory Access}

We begin with the study of different memory access patterns (e.g., sequential and random), read and write operations, and their impacts along with the trail of different data object sizes. 
%
%
Specifically, we develop a comprehensive microbenchmark suite that includes sequential read/write with strided access and random read/write with pointer chasing. By doing so, this microbenchmark captures a broad spectrum of memory access behaviors commonly observed in HPC workloads. For local memory access, we leverage NUMA capabilities by utilizing the \texttt{cpunodebind} and \texttt{membind} provided by the \texttt{numactl} tool. For remote memory access, we employ \texttt{IBV\_WR\_RDMA\_WRITE} and \texttt{IBV\_WR\_RDMA\_READ} operations from the RDMA Verbs library over Ethernet and InfiniBand. We vary the size of data objects by adjusting the buffer allocation size for each memory access operation.





Figure~\ref{fig:motivation1} illustrates the normalized latency (i.e., slowdown) of four memory access patterns – sequential read, sequential write, random read, and random write – when using remote memory over RDMA (both \texttt{Ethernet} and \texttt{InfiniBand}) versus local memory. All results are normalized to the \texttt{Oracle}) case (local memory access).
We have the following observations:

\textbf{(a) Read vs. Write.} In both sequential and random access patterns, remote write operations consistently achieve better performance (lower normalized latency) than read operations. At small data sizes (1–8 KB), the difference is modest (both reads and writes incur only a few microseconds of latency over \texttt{InfiniBand} (e.g., 2-6 µs), making them several tens slower than local memory. However, as the data object size increases, the writes pull ahead significantly. For example, at a 4 MB transfer size, an \texttt{InfiniBand} sequential write completes in 424.46 µs, whereas a sequential read takes 1561 µs (nearly $3.68\times$ longer). A similar trend is seen in random accesses: at 4 MB, an \texttt{InfiniBand} random write finishes in 461.92 µs, versus 1599.7 µs for a random read ({$3.47\times$} slower). 

\summary{These gaps highlight an asymmetry in RDMA operations that one-sided writes can stream large data to the remote node more efficiently, while reads introduce additional round-trip overhead (waiting for the remote memory to be fetched and returned). }
\vspace{3pt}

\textbf{(b) Sequential vs. Random Access.} 
For local memory access, the performance difference between sequential and random reads is not obvious. Only when the data object size is 4 MB does random read exhibit a noticeable slowdown, taking 580 µs compared to 445 µs for sequential read, a 1.3$\times$ difference. In other cases, the gap is negligible.
Similarly, for write operations with data objects smaller than 32 KB, the access pattern (sequential vs. random) has little impact on performance. However, when the data size exceeds 512 KB, the access pattern begins to significantly affect local write performance. For instance, at 4 MB data size, random write to local memory takes 1058 µs, whereas sequential write only takes 557 µs, resulting in a 1.90× difference.
In contrast, for remote memory access (both read and write), the access pattern has almost no impact on performance across all data sizes. 

\summary{For local memory, sequential access benefits from hardware features like CPU prefetching and cache locality. This leads to substantial performance advantages over random access. In contrast, RDMA operations are executed by the NIC using DMA over PCIe. The NIC handles memory requests by pulling tasks from its queue and executing them independently, without CPU-side caching or prefetching. Therefore, sequential and random access patterns in remote memory have similar latencies.}
\vspace{2pt}

\textbf{(c) Different sizes of data objects.} 
(i) For data objects smaller than 4 KB (i.e., smaller than typical OS page sizes), remote memory access latency is over 100$\times$ that of local memory in many cases. This is consistent across all four subfigures in Figure~\ref{fig:motivation1}. These small transfers suffer from fixed RDMA overheads such as connection setup, message metadata processing, and PCIe transactions. Those costs that are negligible for local NUMA memory access but significant for RDMA, particularly over \texttt{Ethernet}. For instance, both sequential and random 1–2 KB reads and writes over \texttt{InfiniBand} or \texttt{Ethernet} exhibit this gap.
(ii) For objects $\ge$ 512 KB, especially in random write scenarios, we observe a surprising trend: InfiniBand-based random remote write can outperform local memory writes. For example, a 512 KB random write completes in 60.4 µs over \texttt{InfiniBand}, better than local (\texttt{Oracle}) write latency. This result is due to the architectural benefit of RDMA NICs (HCAs) bypassing the CPU cache hierarchy, eliminating the penalties of cache eviction and dirty writebacks. In contrast, large local random writes can pollute caches and trigger expensive coherence traffic or writebacks in NUMA systems. Furthermore, on NUMA systems, so-called “local” memory writes typically traverse interconnects (e.g., UPI/QPI), depending on CPU affinity, causing bus contention and congestion. Hence, under certain conditions, remote memory, especially write-heavy and large-object scenarios, can outperform local access.
(iii) Overall, as object size grows, RDMA overheads are amortized, and normalized performance improves. For example, sequential \texttt{InfiniBand} reads jump from a 21.9$\times$ slowdown at 32 KB to only 3.5$\times$ at the size of 4 MB. At that point, remote access becomes bandwidth-bound rather than latency-bound.

\summary{Tiny transfers pay a steep RDMA overhead, but large blocks sent as remote one-sided writes can actually outperform local writes because they bypass CPU cache traffic and coherence delays.}

\subsection{Data Objects in HPC Applications}

\begin{figure*}[!ht]
    \centering
    \begin{subfigure}[t]{0.3\textwidth}
        \captionsetup{skip=-1pt} 
        \includegraphics[width=\textwidth]{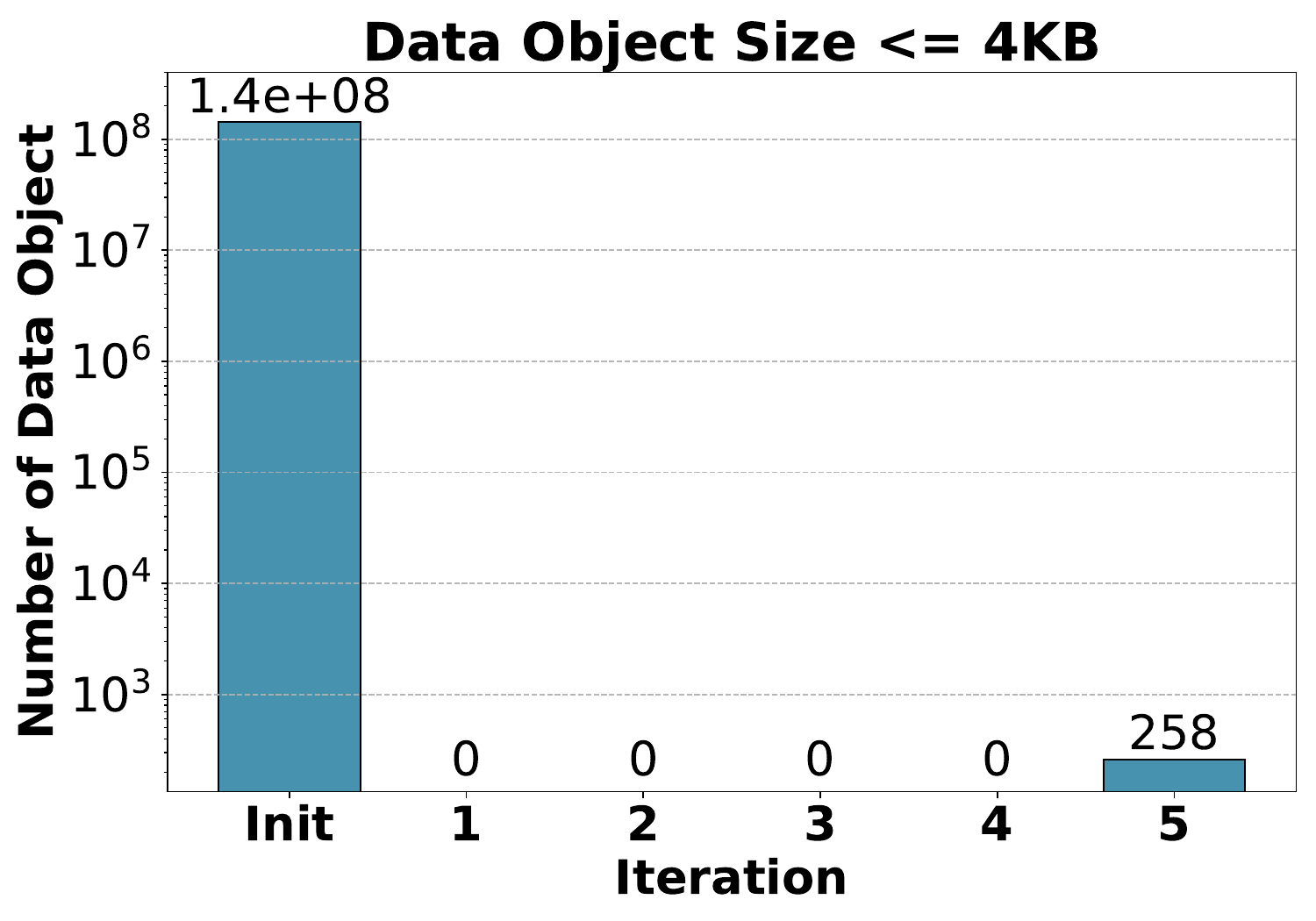}
        \caption{Distribution of data objects that fit in one page.}
        \label{fig:motivation2_a}
    \end{subfigure}
    \hspace{0.01\textwidth}
    \begin{subfigure}[t]{0.3\textwidth}
        \captionsetup{skip=-1pt}
        \includegraphics[width=\textwidth]{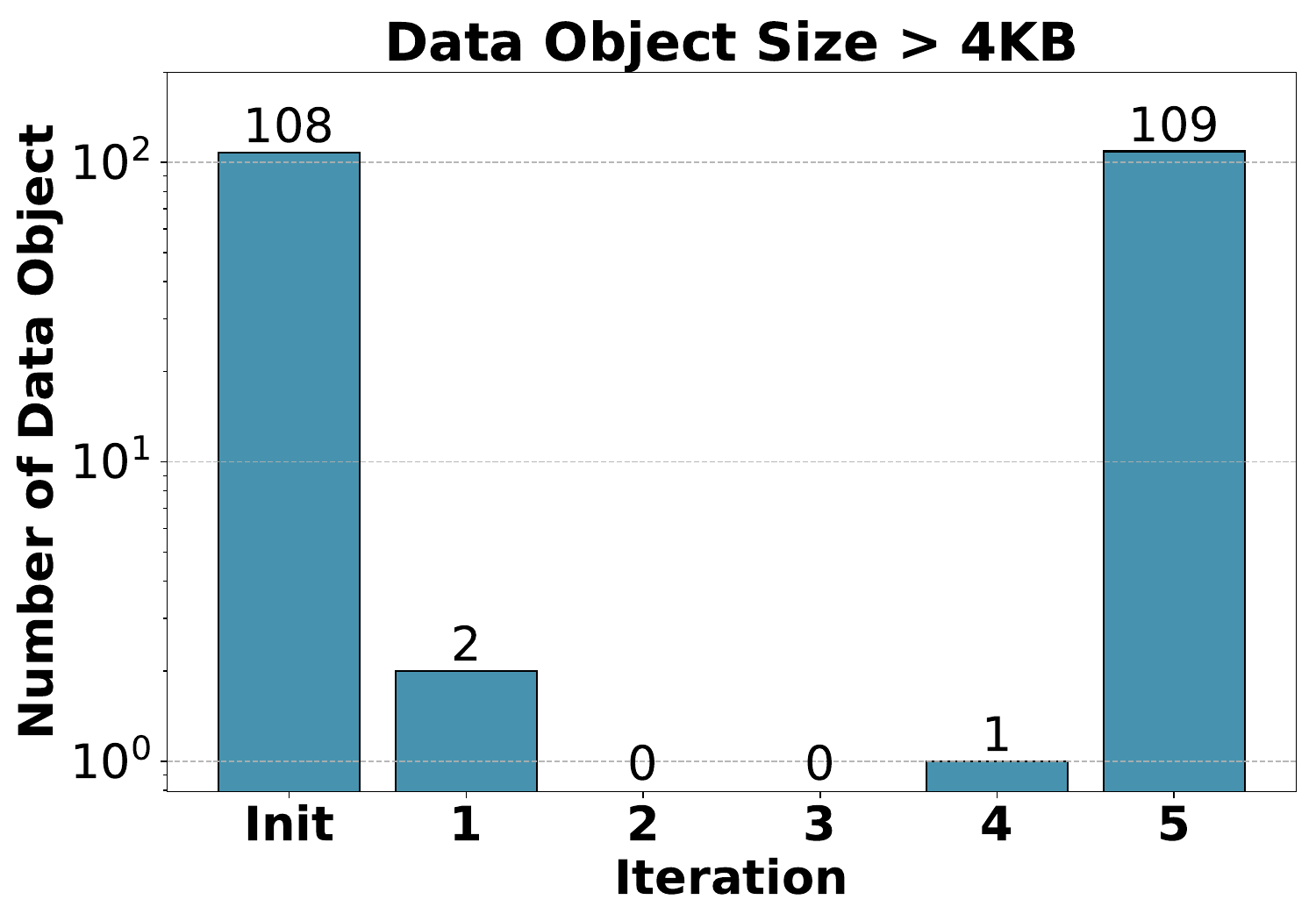}
        \caption{Distribution of data objects larger than one page. }
        \label{fig:motivation2_b}
    \end{subfigure}
    \hspace{0.01\textwidth}
    \begin{subfigure}[t]{0.3\textwidth}
        \captionsetup{skip=-1pt}
        \includegraphics[width=\textwidth]{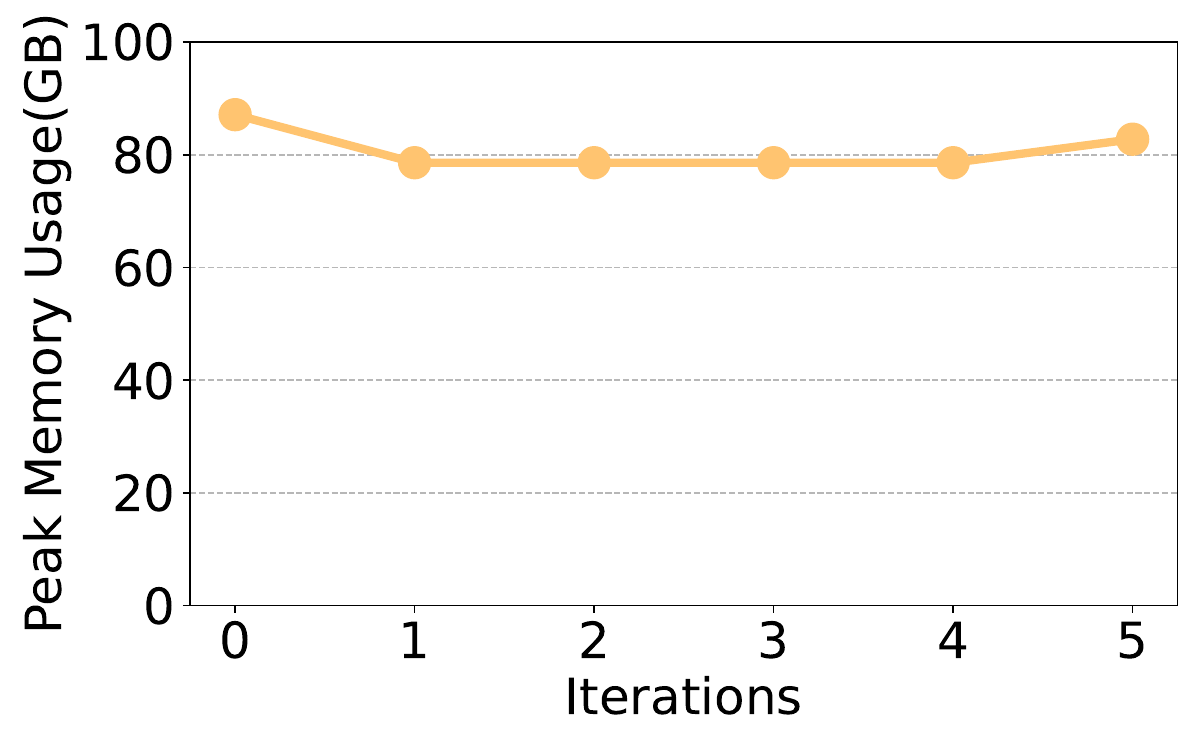}
        \caption{Peak memory consumption across iterations. }
        \label{fig:motivation2_c}
    \end{subfigure}
        \vspace{-5pt}

    \caption{Statistic of data objects in HPC application Laghos.} 
    \label{fig:motivation2}
    \vspace{-5pt}
\end{figure*}



In HPC applications, heap and global data objects often constitute the most important data structures and account for the major memory footprint. To better understand the characteristics of these data objects, we investigate their size and lifetime in the context of Laghos~\cite{laghos, laghos_github}, an HPC application that solves the time-dependent Euler equations. Laghos, like many other HPC applications, is characterized by an iterative structure where a main computation loop dominates the computation time~\cite{cluster20_easycrash}.
We run the Laghos simulation, which reaches a peak memory usage of 88.7 GB during the entire program execution. The execution consists of an initialization phase followed by five simulation iterations under our configuration.
We measure the \textit{lifetime} (i.e., the number of iterations between allocation and deallocation) for all data objects in Laghos. Figure~\ref{fig:motivation2} presents the results. Our observations are as follows:

(1) A limited number of data objects dominate the major memory consumption.We analyze the peak memory usage over the entire program execution, distinguishing between small (i.e., those smaller than 4KB) and large (i.e., those larger than 4KB) data objects. For small data objects , the peak memory usage only reaches 27 MB. In contrast, large data objects  account for a significantly higher peak memory usage of 88.7 GB.
As shown in Figure~\ref{fig:motivation2_a} and~\ref{fig:motivation2_b}, among the 142 million data objects used in Laghos, only 200 data objects are larger than 4KB. However, the peak memory consumption of these large data objects accounts for over 99\% of the total peak memory consumption, as shown in figure~\ref{fig:motivation2_c}. 

(2) A large number of data objects are short-lived (i.e., lifetimes smaller than one iteration) and small-sized (i.e., smaller than 4KB). Figure~\ref{fig:motivation2_a} and~\ref{fig:motivation2_b} shows that 142 million small data objects are created during initialization phase. However, these data objects are deallocated before the next iteration, and only 258 small data objects persist throughout the entire program execution. These objects are mainly intermediate data objects used for computation.

We noticed similar observations consistent across other HPC applications ~\cite{10.1145/3447818.3460356}. 
\textit{We choose data object-level management for HPC applications because managing long-lived, large data objects is more lightweight compared to memory page-level management in disaggregated memory systems.}

\section{Design}
In this section, we introduce \name with its three major components: (1) identify which data objects should be placed in remote memory, (2) how to move data objects between local and remote memory, and (3) how to handle multi-threading.

\subsection{Remote Data Objects Selection} 
Based on our preliminary study on remote memory access performance and data objects in HPC applications (Section~\ref{sec:motivation}), \name employs the following principles to determine which data objects should be placed into remote memory:

(1) Large data objects (i.e., sizes larger than 4KB) are prioritized for placement in remote memory to maximize local memory savings. \name dynamically ranks data objects by size in descending order during runtime, based on the information available at allocation time. When local memory capacity is insufficient, \name moves the largest data objects to remote memory first.

(2) Among large data objects of the same size, \name prioritizes those with the lowest number of accesses for placement in remote memory. 
Frequent remote accesses, especially read-after-write patterns, can lead to concurrency overhead and make remote memory data structures less efficient than distributed alternatives.

(3) Among large data objects with the same size and same access frequency, \name prioritizes those with more writes for placement in remote memory. This leverages the observation that remote memory prefers write than read.

While \name primarily focuses on placing large, long-lived heap data objects in remote memory, it also addresses the need to efficiently manage smaller, short-lived data objects. For these data objects, \name leverages atomic operations on remote memory supported by RDMA~\cite{rdma}. By using atomic operations, \name enables efficient and concurrent access to these data objects in remote memory, ensuring data integrity and consistency.

To guarantee the correct sequencing and consistency of remote memory accesses, \name relies on the memory fabric's ability to enforce ordering constraints. The memory fabric ensures that memory operations issued before a memory barrier (or fence) complete before any operations issued after the barrier. This ordering functionality can be implemented using request completion queues in the memory fabric interface. By leveraging these ordering constraints, \name maintains the correct order of remote memory accesses, even in the presence of concurrent operations from multiple processors.

\subsection{Memory Management of \name}

\begin{figure*}[t]   
    \centering
    \includegraphics[width=\textwidth]{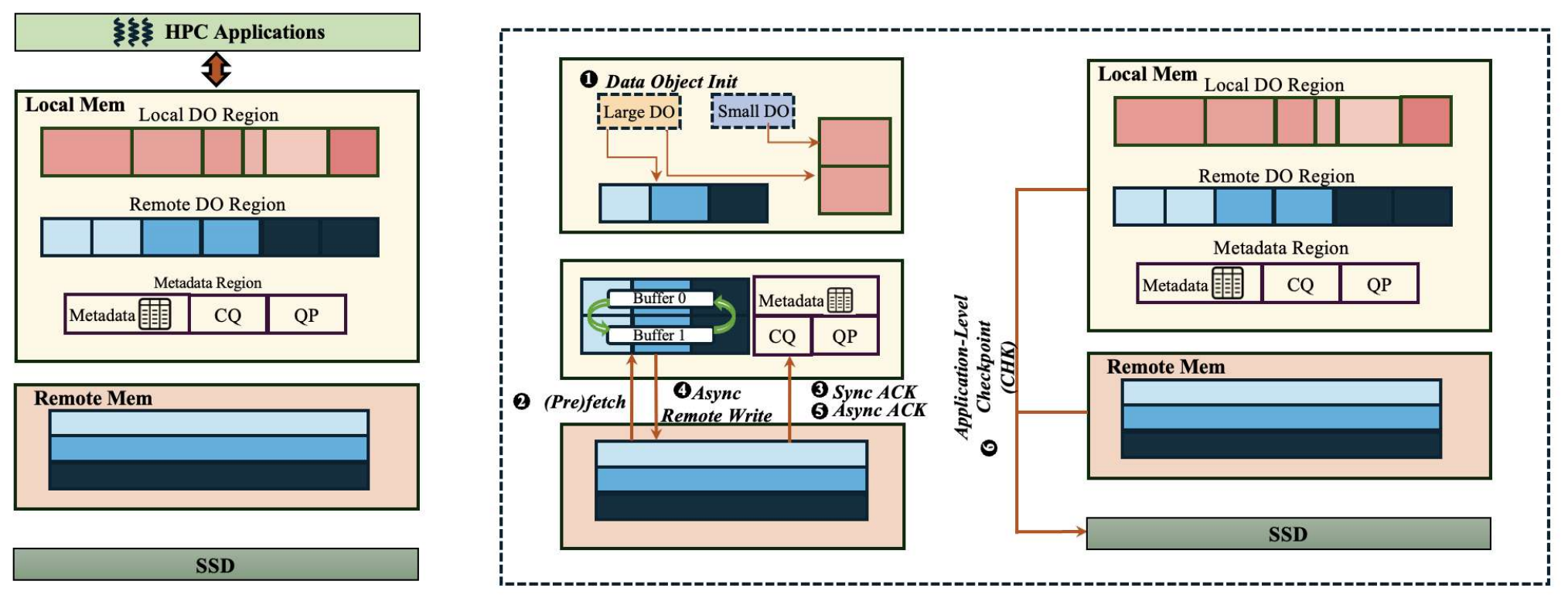}
    \caption{The workflow of \name. ``DO'' is an abbreviation for ``data objects'', represented by the colored dashed boxes.}
    \label{fig:workflow}
\end{figure*}

We introduce the detailed memory management strategy of \name. Figure~\ref{fig:workflow} illustrates the workflow of \name. Specifically, \name separates the local memory space into three parts: the local data object region, the remote data object region, and the metadata region. The local data object region stores data objects that are frequently accessed or have small sizes, to avoid frequent data movement. 
The remote data object region is an RDMA registered memory space and acts as a software-managed cache for remote memory.
The metadata region is used to store metadata such as queue pairs (QP) and completion queues (CQ) required for RDMA communication, as well as lightweight metadata for \name that records which data objects (i.e., special virtual address and offsets) are cached in local memory and their status. 
We demonstrate an example on how \name handles different type of data objects.

\textbf{Data object initialization.}
Initially, all data objects are allocated in local memory, as shown in step \ding{182} in figure~\ref{fig:workflow}. 
\name intercepts memory allocation APIs to gain control over the allocation process. 
For small data objects that fit within the local memory capacity, \name allocates them directly in the local data object region, ensuring fast access to frequently used data. 
When a data object exceeds the capacity of the available local memory but smaller than the local data object region, \name first migrates existing data objects from local to remote memory before performing local allocation. 
When allocating a data object larger than local data object region, \name directly returns the remote memory address and performs the allocation in the remote memory.
This approach ensures efficient utilization of local memory while seamlessly managing large data objects in remote memory.

\textbf{Remote read with dual buffer.}
In \name, all remote data objects are assigned a special virtual address that redirects access to the metadata table. When an application attempts to access a data object that is not present in local memory, \name triggers an on-demand fetching mechanism to retrieve the required data object from remote memory and store it in the remote data object buffer (shown as step \ding{183} in figure~\ref{fig:workflow}). 
Simultaneously, the metadata table is updated accordingly to reflect the new location of the data object.
\name employs an access barrier (step \ding{184}) to guarantee that the fetched data object is fully accessible in the local memory before the application proceeds with its computation, preventing any potential data inconsistencies or race conditions. \name also supports indirect memory accesses, though at a higher cost. For instance, an access pattern like A[B[i]] may incur two remote memory accesses if both A and B reside on remote memory, due to the additional level of indirection.


If the size of the remote data object exceeds the capacity of the remote data object region, \name prefetches the largest possible portion 
of the data object that can fit within the available space. This ensures optimal utilization of local memory resources while still providing access to the most critical parts of the data object.

To further mitigate the high overhead associated with on-demand reads, \name implements a dual buffer design in the remote data object region, taking advantage of the iterative nature commonly found in HPC applications. By proactively prefetching data objects required for the next few iterations into the idle buffer, \name effectively overlaps most of the remote read overhead with ongoing computation. Through clever manipulation of buffer pointers to alternate between the two buffers, \name minimizes the performance impact of remote memory access in iterative scenarios, allowing the application to continue executing with minimal delays.

\textbf{Asynchronous remote memory write.}
When the local data object region reaches its capacity, \name demotes data objects stored in the remote memory region to remote memory and updates the metadata table simultaneously (step \ding{185} in Figure~\ref{fig:workflow}). To minimize the impact of remote memory writes on application performance, \name employs an asynchronous writing mechanism (step \ding{186}). Specifically, \name does not wait for the acknowledgment of write completion but continues with the computation.

\name supports a concurrent access model that enables multiple compute nodes can access and modify the data objects. The model specifies whether one or many compute nodes can write to the data objects and whether one or many compute nodes can read them by adding a write lock on remote memory. This flexibility enables \name to adapt to different application requirements and optimize performance based on access patterns.

The asynchronous writing mechanism in \name explores the trade-off between fast remote reads and fast remote writes. By prioritizing the performance of remote writes, \name aims to minimize the overhead of data demotion and ensure that the application can continue executing without significant delays. However, this optimization may come at the cost of slower remote reads, as the data objects may not be immediately available in the local memory.

\name's design choice aligns with the nature of MPI (Message Passing Interface), which is commonly used in HPC applications. MPI typically partitions the computation tasks among compute nodes, with each node responsible for a specific portion of the data. Due to the expensiveness of communication in MPI, most MPI programs do not contain large shared data objects, which avoids concurrent accesses to remote memory in the context of memory disaggregation. Therefore, \name chooses asynchronous remote write but synchronous remote read to maximize performance.


\textbf{Reliability and failure handling. }
In a disaggregated memory environment, computation nodes and memory nodes can fail independently. To ensure the reliability and fault tolerance of the system, \name incorporates a checkpoint-based recovery mechanism. As shown in step \ding{187} of Figure~\ref{fig:workflow}, \name periodically saves application-level checkpoints (CHK) to persistent storage, such as SSDs.

\name employs an asynchronous checkpointing approach to minimize the impact on application performance. It captures the state of both local memory and remote memory separately, allowing the application to continue execution while the checkpoints are being saved. This asynchronous nature of checkpointing ensures that the application's progress is not stalled during the checkpoint process. 
To maintain consistency between the local and remote memory checkpoints, \name leverages the metadata table. The metadata table keeps track of the mapping between local memory and remote memory addresses, as well as the status of each data object. During the checkpointing process, \name identifies the data objects that have been modified in the local memory since the last checkpoint. It then updates the corresponding entries in the remote memory checkpoint with the latest values from the local memory checkpoint. This selective update mechanism ensures that the remote memory checkpoint reflects the most recent state of the application. 

When there is a failure, \name initiates a recovery process by loading the checkpoint back to remote memory. 
The metadata table is also restored from the checkpoint, ensuring that the mapping between local and remote memory remains accurate for all data objects.

\begin{table*}[htbp]
\centering
\caption{Summary of evaluated workloads and their memory characteristics}
\label{table:workloads}
\small 
\setlength{\tabcolsep}{4pt} 
\begin{tabular}{l l c c c c}
\toprule
\textbf{Workload} & \textbf{Characteristics} & \textbf{Total Memory (GB)} & \textbf{Read/Write Ratio} & \textbf{Data Objects} & \textbf{Remote Memory (GB)} \\
\midrule
NPB:CG & Irregular, non-sequential access & 8.6 & 1:1 & \texttt{a} & 5.4 \\
NPB:MG & Hierarchical, semi-regular access & 26.5 & 9:8 & \texttt{u,v,r} & 26.4 \\
NPB:FT & Non-sequential, multi-dimensional access & 80.0 & 11:7 & \texttt{twiddle,u\_0,u\_1} & 80.0 \\
NPB:BT & Intra-block, irregular inter-block access & 10.7 & 5:3 & \texttt{u,forcing,rhs} & 7.6 \\
NPB:LU & Non-uniform access & 8.8 & 15:8 & \texttt{u,rsd,frct} & 7.6 \\
NPB:IS & Sequential, parallel access & 32.3 & 1:1 & \texttt{key\_array,key\_buf2} & 32.0 \\
XSBench & Random access, lookup intensive & 5.5 & 1:1 & \texttt{index\_grid} & 5.1 \\
miniAMR & Hierarchical access, irregular patterns & 32.2 & 11:9 & \texttt{blocks} & 30.9 \\
\bottomrule
\end{tabular}
\end{table*}

\subsection{Handling Multi-threading within One Compute Node}

HPC applications commonly leverage OpenMP for multi-threaded execution within compute nodes.
Integrating multi-threading with remote memory access introduces two key challenges. First, when shared data objects reside in remote memory, OpenMP threads must coordinate their remote memory operations to maintain data consistency. 
Second, concurrent RDMA operations from multiple threads can lead to resource contention, as each thread requires dedicated RDMA resources for remote memory access~\cite{asplos24_smart}. 

To maximize performance for the common case of non-shared (thread private) data objects, \name adopts a per-thread local buffer design. The available local buffer space is equally partitioned among OpenMP threads, with each thread maintaining exclusive access to its buffer partition. 
This design aligns naturally with OpenMP's private variable semantics, where threads operate on independent data. 
Each thread uses its dedicated buffer space for remote data operations, eliminating synchronization overhead and enabling efficient parallel access.

To manage RDMA resource contention in multi-threaded scenarios, \name implements a two-level scheduling design. At the lower level, threads are organized into clusters to share RDMA resources, where each cluster receives dedicated RDMA resources to prevent system-wide contention. 
At the upper level, within each cluster, threads coordinate their RDMA operations through a shared scheduling queue while maintaining their private local buffers. 
This hierarchical approach balances the performance benefits of per-thread buffers with efficient RDMA resource utilization when numerous threads perform remote memory operations simultaneously. 

While per-thread buffers efficiently handle private data access, coordinating remote memory operations for shared data objects requires additional synchronization. \name implements a fine-grained locking mechanism where each shared object maintains its own lock in local memory, enabling concurrent access to different shared objects while ensuring data consistency. 
While this locking mechanism introduces additional overhead for shared object access, this limitation aligns with typical OpenMP programming patterns where shared objects are minimized to avoid performance degradation in parallel regions. Most OpenMP applications already limit shared data access to maximize parallel efficiency, making this a practical tradeoff for our design. 


\section{Implementation}
\name is implemented using one-sided RDMA operations in C and C++. The implementation provides APIs for remote memory accesses, including remote allocation/deallocation,  remote read/write, and synchronous/asynchronous acknowledgment (ACK). 

To reduce the overhead of remote memory access, \name implements a dual buffer design in the remote data object region. This implementation relaxes the synchronization requirement for remote reads by modifying the strict barrier from immediately after the remote read operation to just before the computation that uses the remotely read data. By deferring the synchronization until the data is actually needed for computation, \name minimizes the waiting time associated with remote memory accesses.



\section{Evaluation}
\begin{figure*}[!htbp]
\captionsetup[subfigure]{position=above}
    \centering
    \begin{subfigure}{0.24\textwidth}
        \centering
        \captionsetup{skip=-0.2em}
        \caption{CG}
        \includegraphics[width=\linewidth]{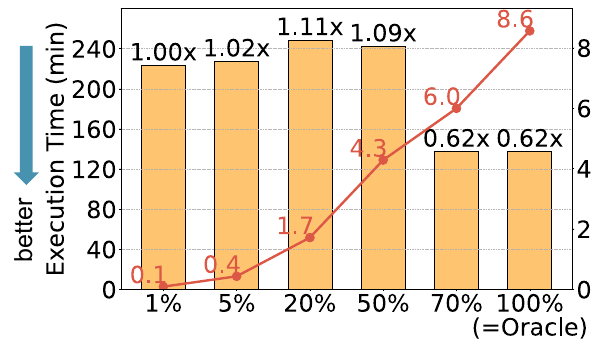}
        \label{fig:cg}
    \end{subfigure}
    \begin{subfigure}{0.23\textwidth}
        \centering
        \captionsetup{skip=-0.2em}
        \caption{MG}
        \includegraphics[width=\linewidth]{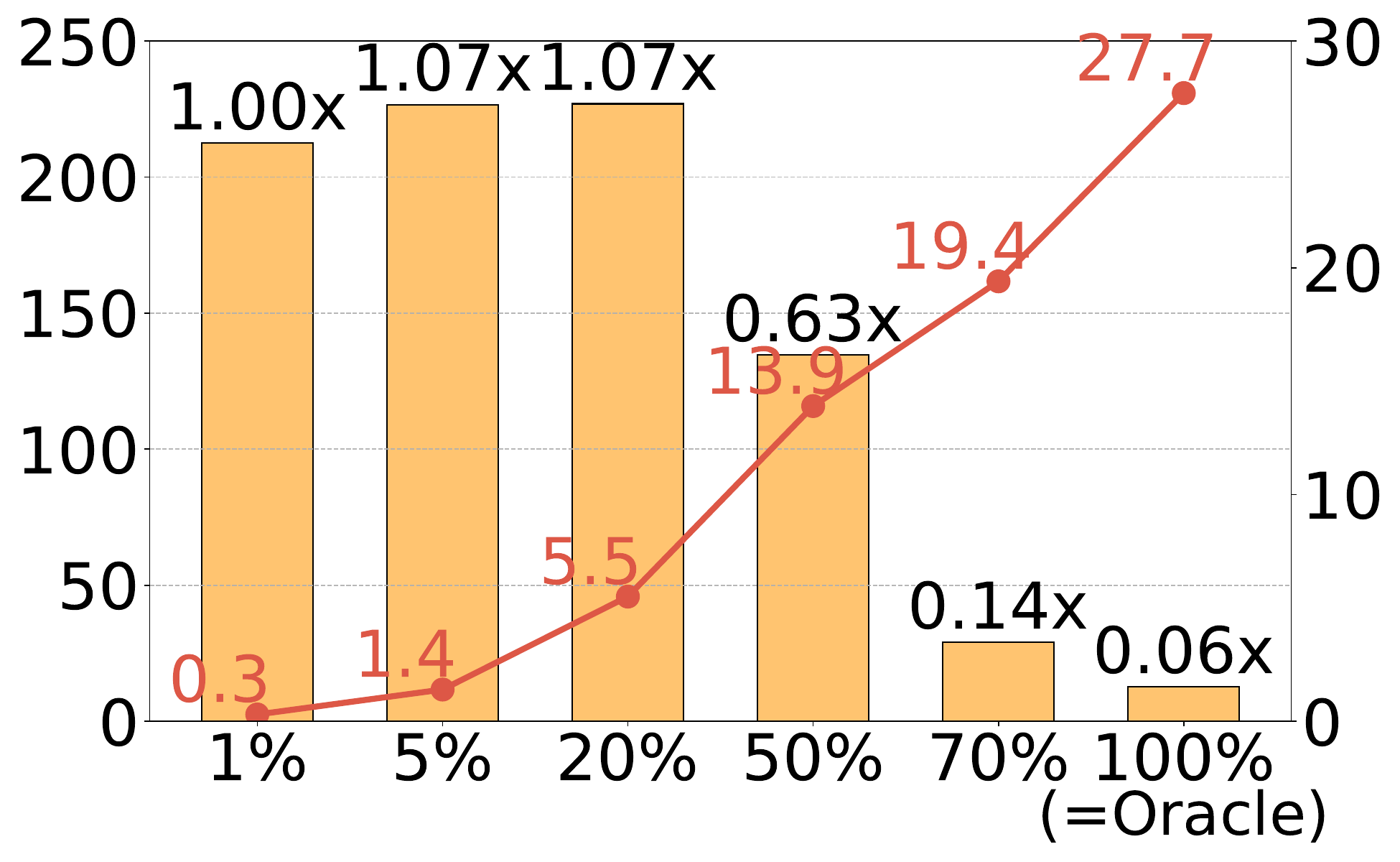}
        \label{fig:mg}
    \end{subfigure}
    \begin{subfigure}{0.23\textwidth}
        \centering
        \captionsetup{skip=-0.2em}
        \caption{BT}
        \includegraphics[width=\linewidth]{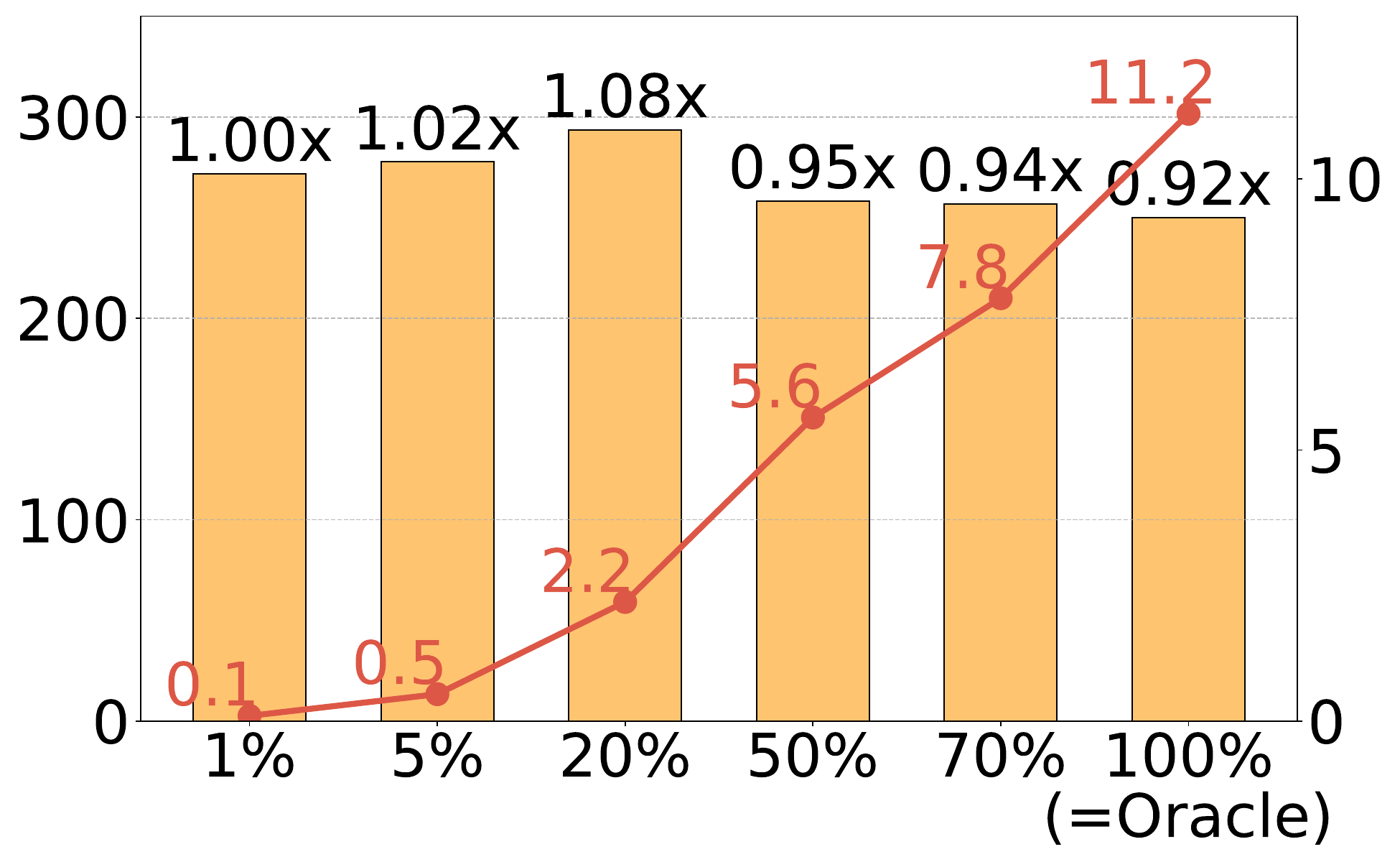}
        \label{fig:bt}
    \end{subfigure}
    \begin{subfigure}{0.23\textwidth}
        \centering
        \captionsetup{skip=-0.2em}
        \caption{FT}
        \includegraphics[width=\linewidth]{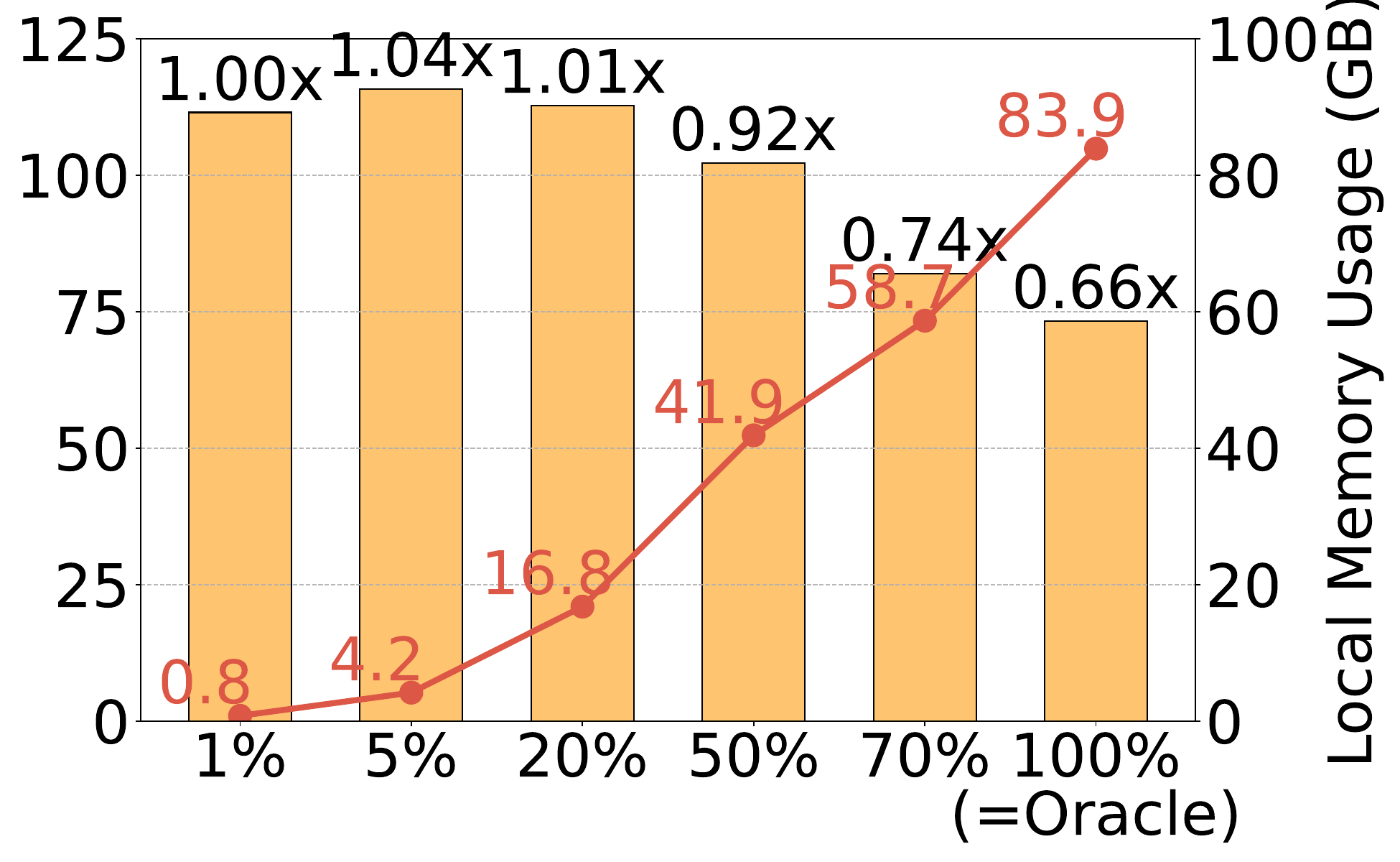}
        \label{fig:ft}
        \label{fig}
    \end{subfigure}

    \centering
    \begin{subfigure}{0.24\textwidth}
        \centering
        \captionsetup{skip=-0.2em}
        \caption{LU}
        \includegraphics[width=\linewidth]{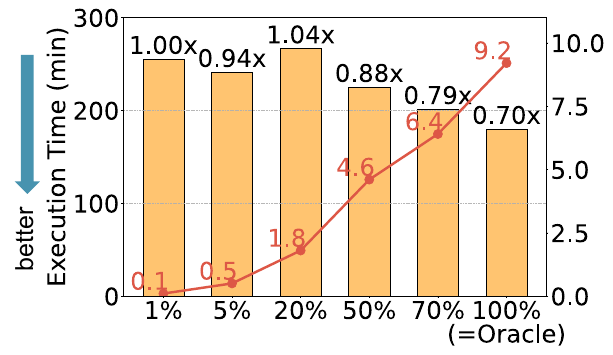}
        \label{fig:lu}
    \end{subfigure}
    \begin{subfigure}{0.23\textwidth}
        \centering
        \captionsetup{skip=-0.2em}
         \caption{IS}
        \includegraphics[width=\linewidth]{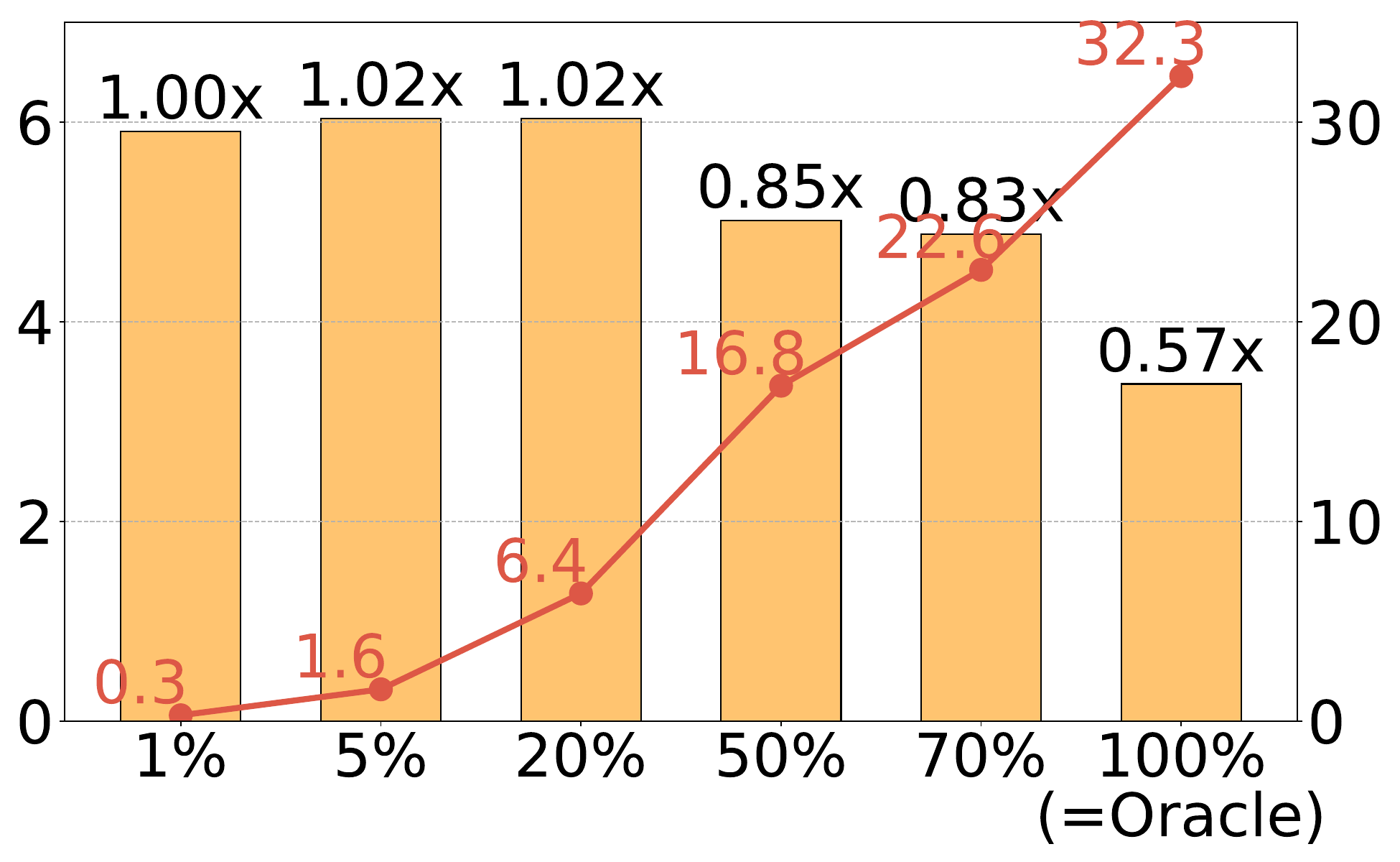}
        \label{fig:ft}
    \end{subfigure}
    \begin{subfigure}{0.23\textwidth}
        \centering
        \captionsetup{skip=-0.2em}
        \caption{XSBench}
        \includegraphics[width=\linewidth]{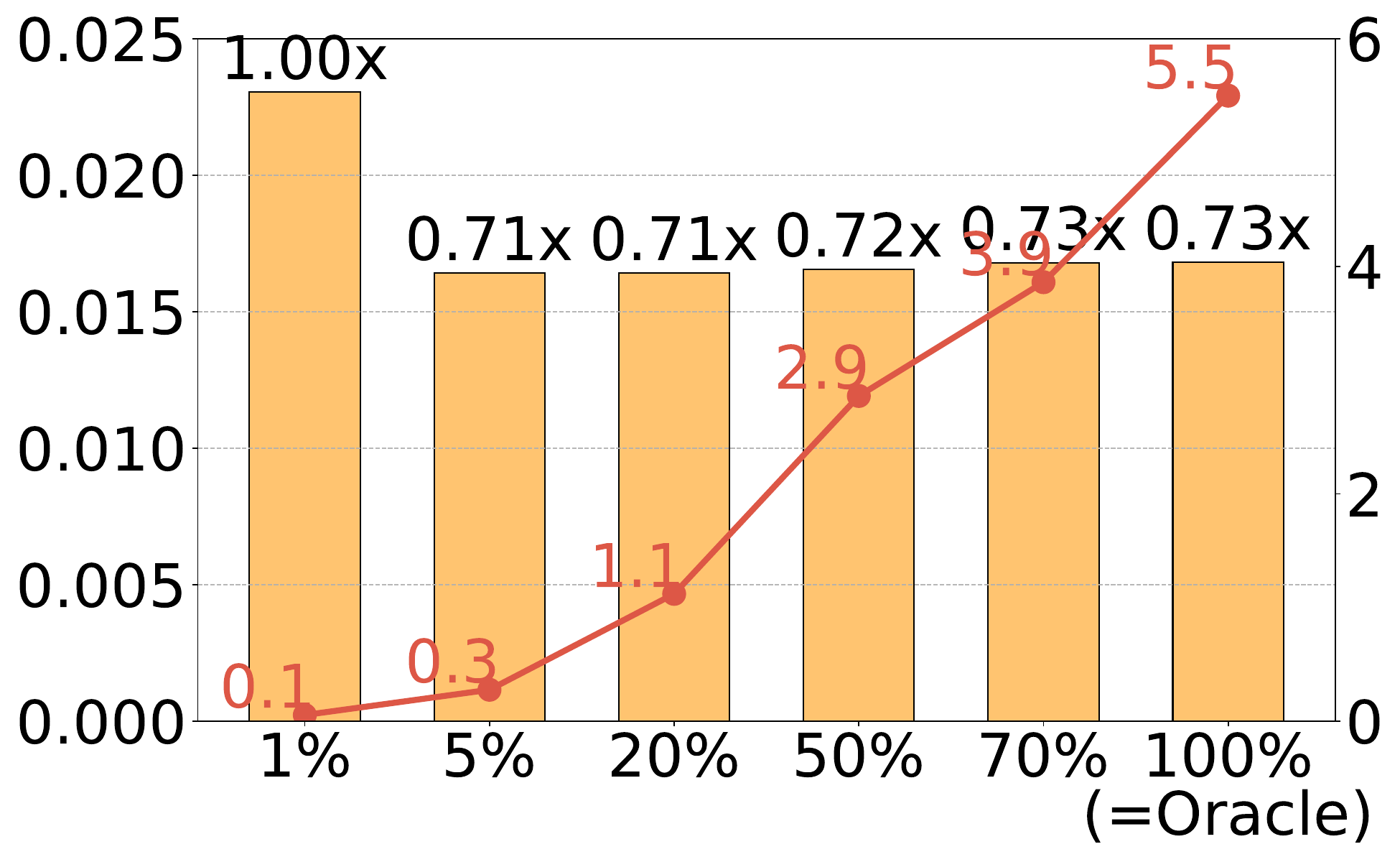}
        \label{fig:ft}
    \end{subfigure}
    \begin{subfigure}{0.23\textwidth}
        \centering
        \captionsetup{skip=-0.2em}
        \caption{miniAMR}
        \includegraphics[width=\linewidth]{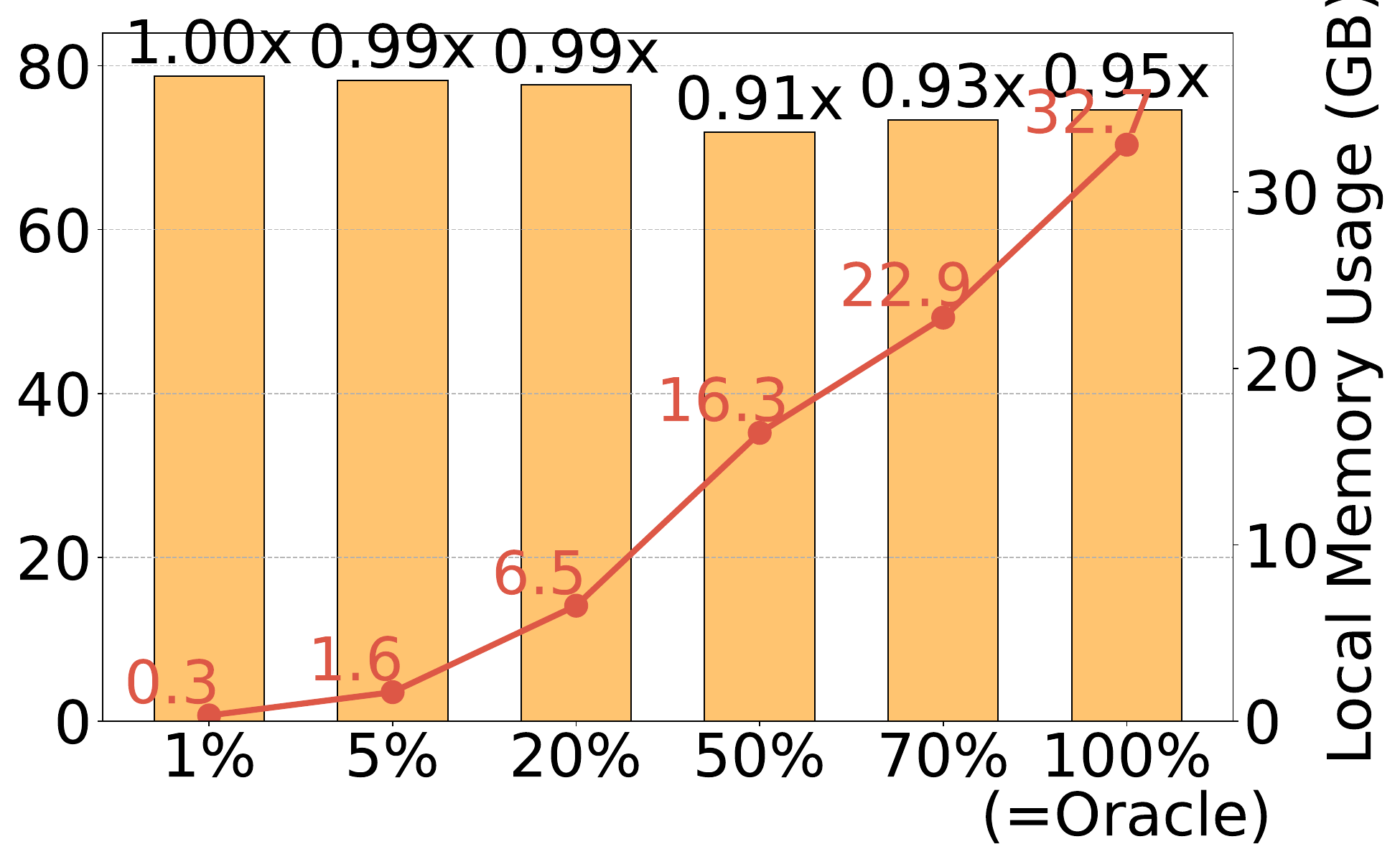}
        \label{fig:ft}
    \end{subfigure}
    \vspace{-10pt}
    \caption{Execution time and local memory capacity across various workloads under different local memory configurations. Each subfigure's x-axis represents the size of the local data object region relative to the application’s peak memory consumption.
    The bars indicate the execution time for each memory configuration, and the red line marks the local memory capacity available for each configuration.
    }
    \vspace{-5pt}
    \label{fig:over_Pef}
\end{figure*}

In evaluation, we use the same environmental settings as Section ~\ref{sec:motivation} with a diverse spectrum of HPC applications. 
Table~\ref{table:workloads} lists the eight workloads evaluated in \name, comprising six from the NAS Parallel Benchmark (NPB)\cite{bailey1991parallel} and two real-world HPC applications: XSBench and miniAMR from the ECP benchmark\cite{messina2017exascale}. The selected workloads encompass diverse memory access patterns and computational intensities, providing a comprehensive assessment of \name’s efficiency.

The NPB workloads include (i)\textit{ CG (Conjugate Gradient)}, which involves irregular memory access in sparse matrix operations; (ii) \textit{MG (Multi-Grid)}, featuring hierarchical grid structures with varied communication patterns; (iii) \textit{FT (discrete 3D Fast Fourier Transform)}, characterized by multi-dimensional data access and all-to-all communication; (iv) \textit{BT (Block Tri-diagonal Solver)}, with predictable intra-block and irregular inter-block memory access; (v) \textit{LU (Lower-Upper Gauss-Seidel Solver)}, demonstrating non-uniform and structured memory access; and (vi) \textit{IS (Integer Sort)}, involving parallel, sequential memory operations. 

The two real-world applications further diversifies the evaluation, as \textit{XSBench} shows random, lookup-intensive access patterns based on Monte Carlo particle transport, and \textit{miniAMR} exhibits hierarchical, adaptive memory access reflecting dynamic simulation behaviors. This diversity ensures a comprehensive evaluation of \name across varied scenarios, demonstrating its capability to optimize memory access and computational efficiency in HPC environments.

\subsection{Overall Performance}
\label{sec:overall_performance}

To evaluate the effectiveness of \name, we measure two key metrics: (i) the \textit{execution time} of the tested workloads, which includes both computation time and all of the data movement overhead at the local node, as well as (ii) the \textit{peak memory usage} of the local memory.

We use \texttt{Oracle} (where
all memory accesses are local) as the baseline.
Here, we vary the space of the registered memory 
at the compute node, which comprises the total size of the remote data object region and the metadata region. 
We configure the local memory space as proportions of the peak memory usage observed in the \texttt{Oracle}, with configurations of 1\%, 5\%, 20\%, 50\%, 70\%, and 100\%. We noticed that the size of local data objects is negligible compared to the overall peak memory consumption.

By varying the registered memory sizes, we can assess the impact of limited local memory on the execution time and analyze the data movement overhead introduced when different proportions of the peak memory usage are available locally. This approach allows us to evaluate the performance of \name under various memory constraints and understand the trade-offs between local memory usage and data movement overhead.
Figure~\ref{fig:over_Pef} illustrates the overall efficiency of \name.

\subsubsection{Execution Time Analysis}

The bars represent the execution time for each workload as the memory size varies. When local memory is highly constrained (e.g., at 1\% and 5\% of the peak memory usage), execution times increase significantly across all workloads due to frequent data exchanges with remote memory, resulting in higher overhead.
The performance degradation observed when increasing the local memory region from very small values (e.g., 1\% to 5\%) is due to the limited size of data that can be transferred in each RDMA operation bounded by the available local memory. Although increasing the local memory slightly allows for larger chunks per transfer, it does not significantly reduce the total number of RDMA operations required for large data objects. As a result, the communication overhead remains high and the performance gain is minimal. Substantial performance improvements are only observed when the local memory region is large enough (e.g., >50\%), allowing significantly fewer and more efficient RDMA transfers.
At the 50\% memory configuration, many workloads show substantial improvement as \name efficiently overlaps computation with data movement. However, in cases like MG and IS (Figures~\ref{fig:over_Pef}(b) and (f)), the intrinsic memory access patterns result in less pronounced gains. When local memory is increased to 70\% or higher, most applications' execution times stabilize and approach the \texttt{Oracle} baseline, demonstrating \name{}’s ability to manage memory efficiently.

\subsubsection{Memory Consumption Analysis}
The red line marks the local memory capacity available
for each configuration. The available local memory closely follows the memory constraints under different local memory configurations. \name's on-demand fetching mechanism to retrieve the required data object from remote memory almost fully utilizes the local memory. However, when the local memory configuration is increased, the performance could be limited by RDMA’s fixed data transfer size per operation (typically 1GB or 2GB).
As a result, multiple transfers may still be required, even when larger local memory is available, leading to potential overhead and longer waiting times. This shows that simply expanding the local memory does not necessarily maximize application performance.

\begin{figure}[!htbp]
    \centering
    \captionsetup[subfigure]{position=above}
    \begin{subfigure}{0.48\linewidth}
        \centering
        \captionsetup{skip=-0.2em}
        \caption{CG (local mem=0.09GB)}
        \includegraphics[width=0.9\linewidth]{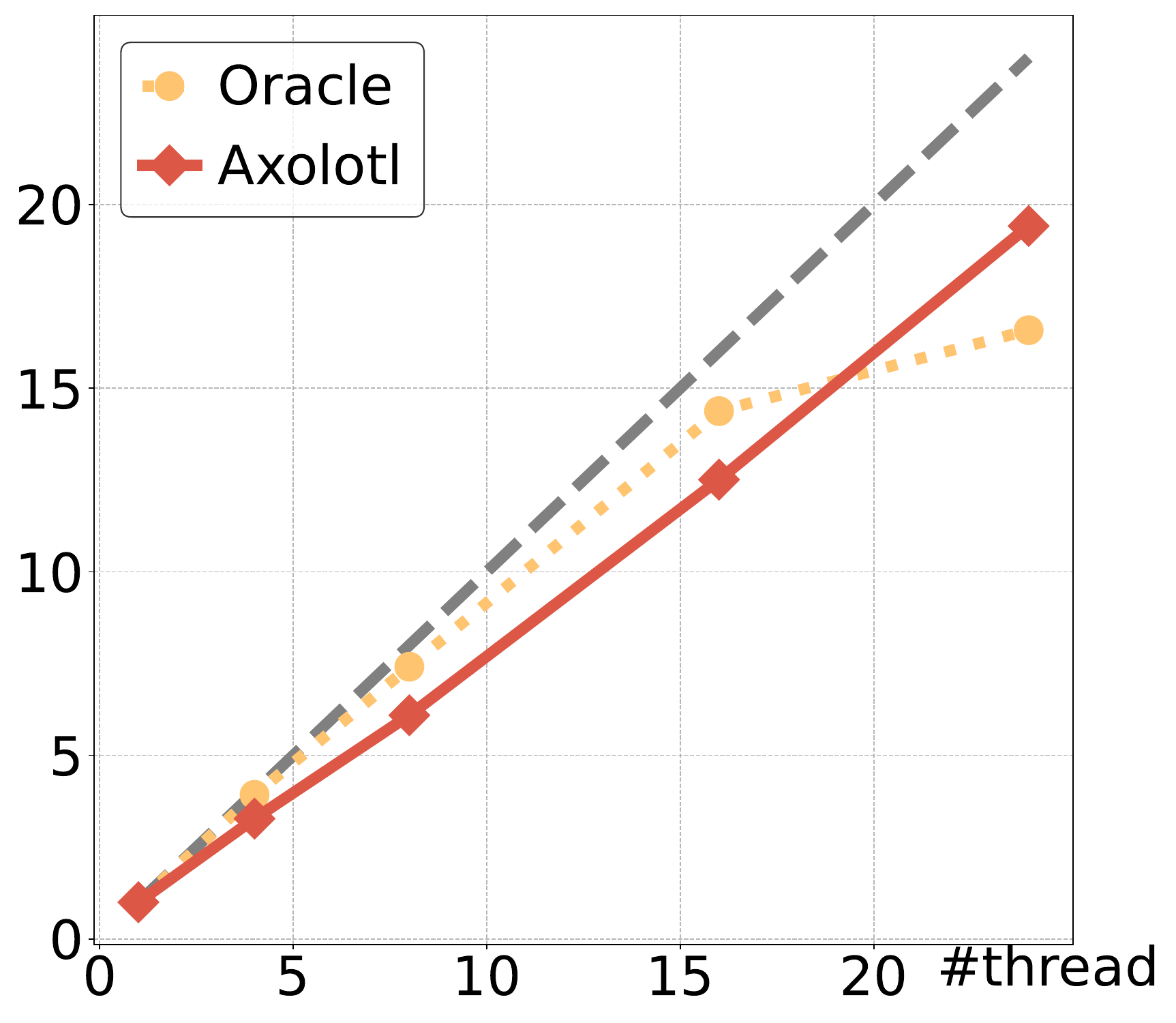}
        \label{fig:cg}
    \end{subfigure}
    \hfill
    \begin{subfigure}{0.48\linewidth}
        \centering
        \captionsetup{skip=-0.2em}
        \caption{MG (local mem=18.54GB)}
        \includegraphics[width=0.9\linewidth]{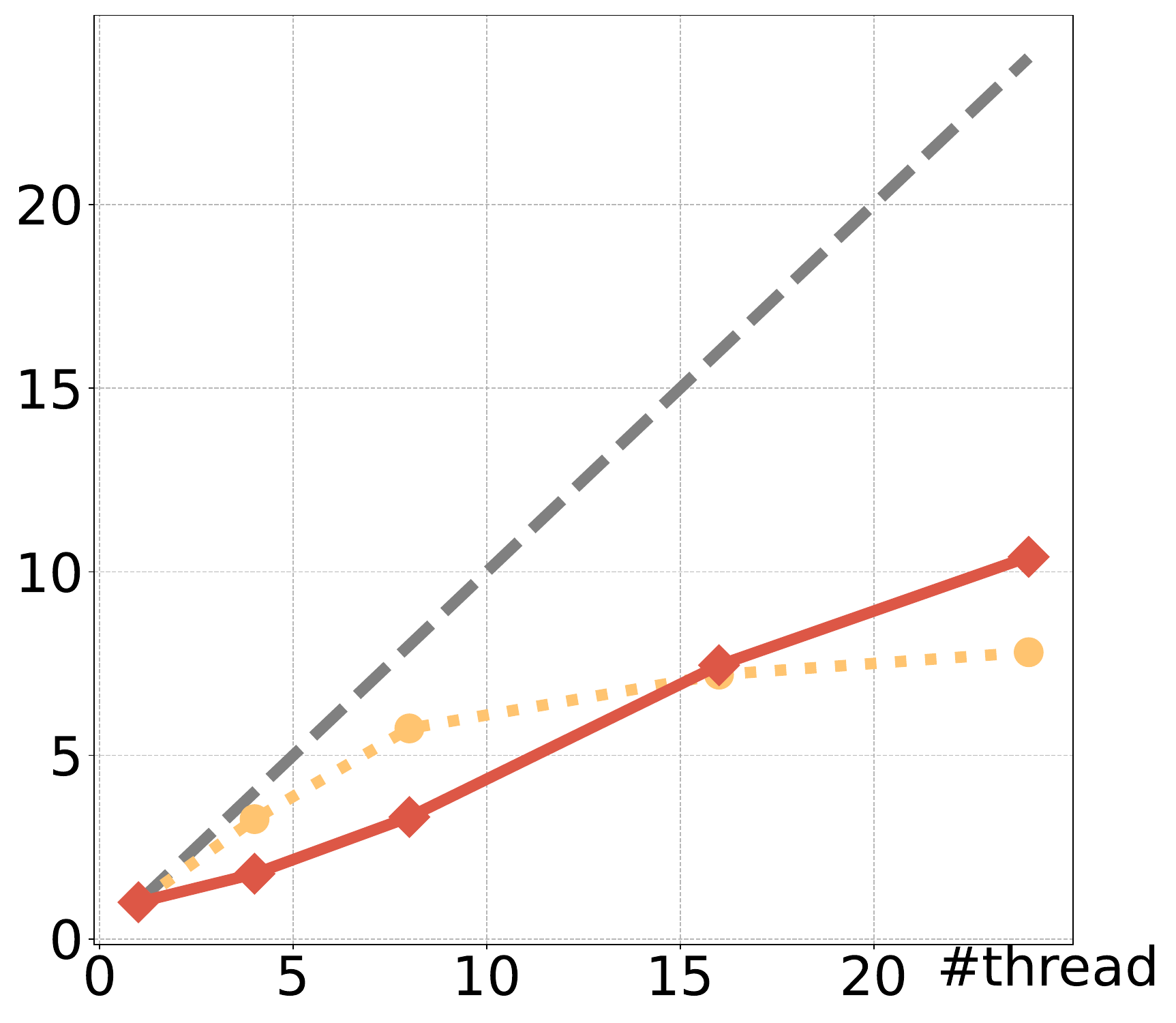}
        \label{fig:mg}
    \end{subfigure}

    \vspace{0.5em} 

    \begin{subfigure}{0.48\linewidth}
        \centering
        \captionsetup{skip=-0.2em}
        \caption{BT (local mem=0.11GB)}
        \includegraphics[width=0.9\linewidth]{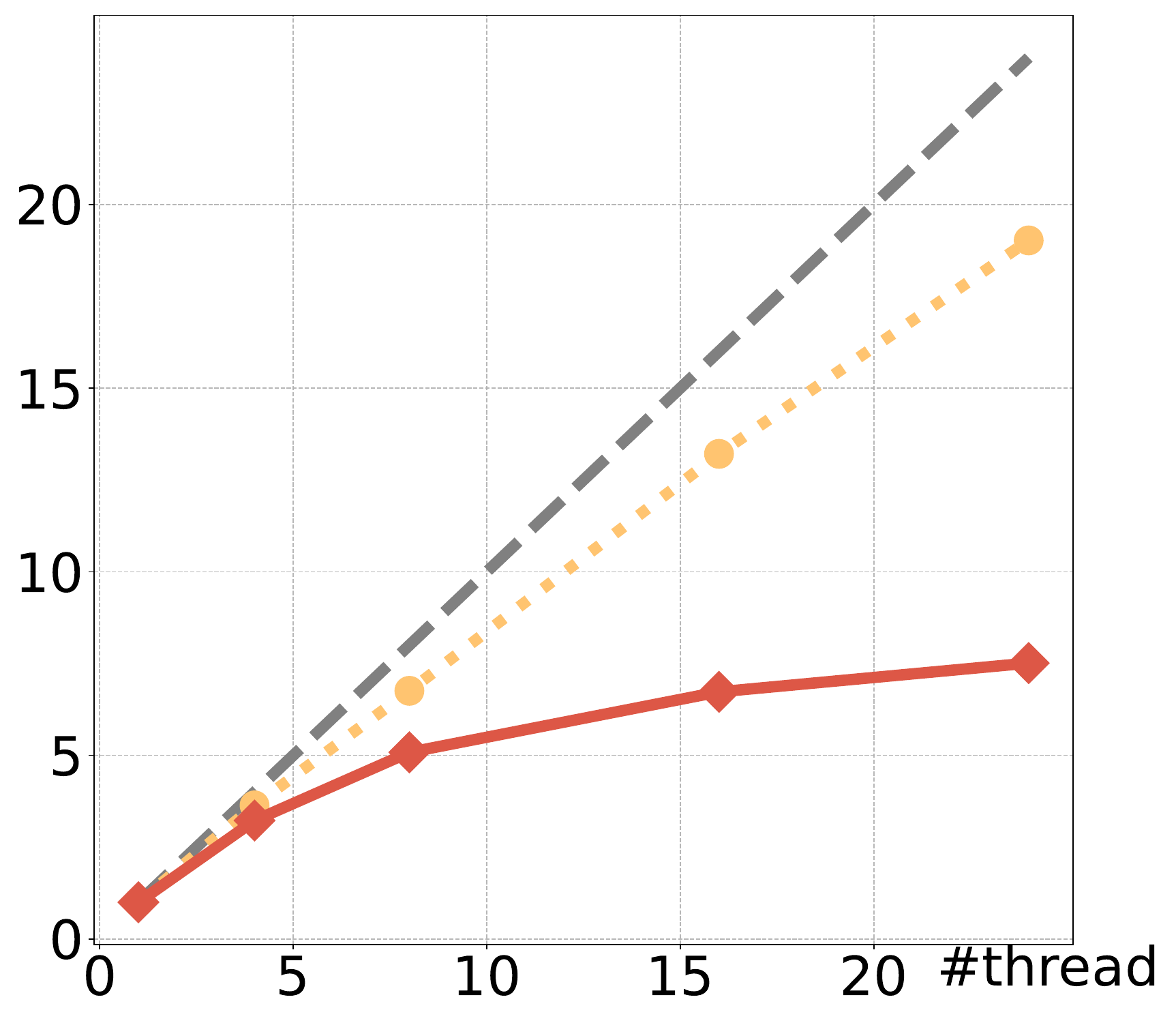}
        \label{fig:bt}
    \end{subfigure}
    \hfill
    \begin{subfigure}{0.48\linewidth}
        \centering
        \captionsetup{skip=-0.2em}
        \caption{FT (local mem=40.00GB)}
        \includegraphics[width=0.9\linewidth]{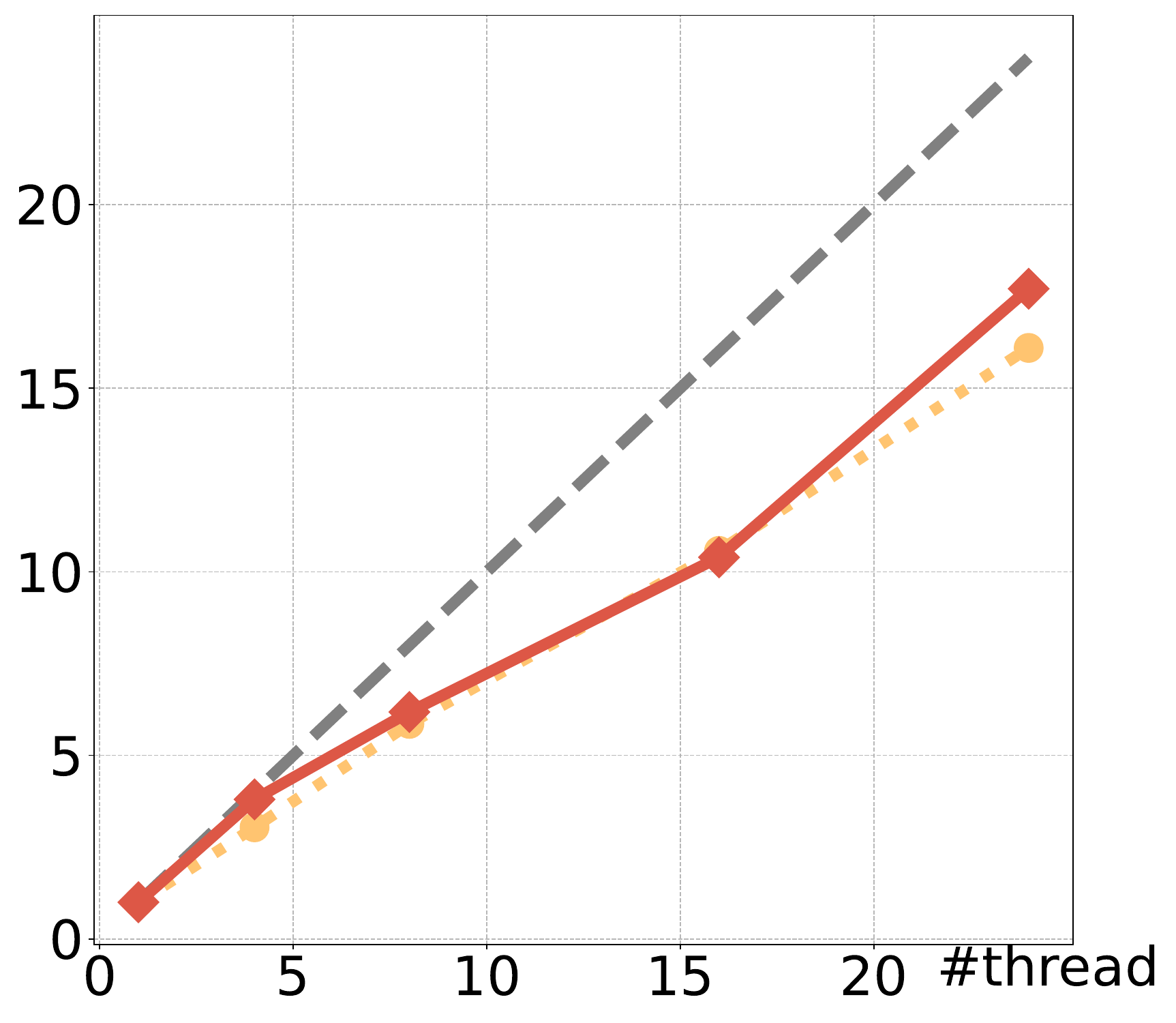}
        \label{fig:ft}
    \end{subfigure}

    \vspace{0.5em} 

    \begin{subfigure}{0.48\linewidth}
        \centering
        \captionsetup{skip=-0.2em}
        \caption{LU (local mem=4.50GB)}
        \includegraphics[width=0.9\linewidth]{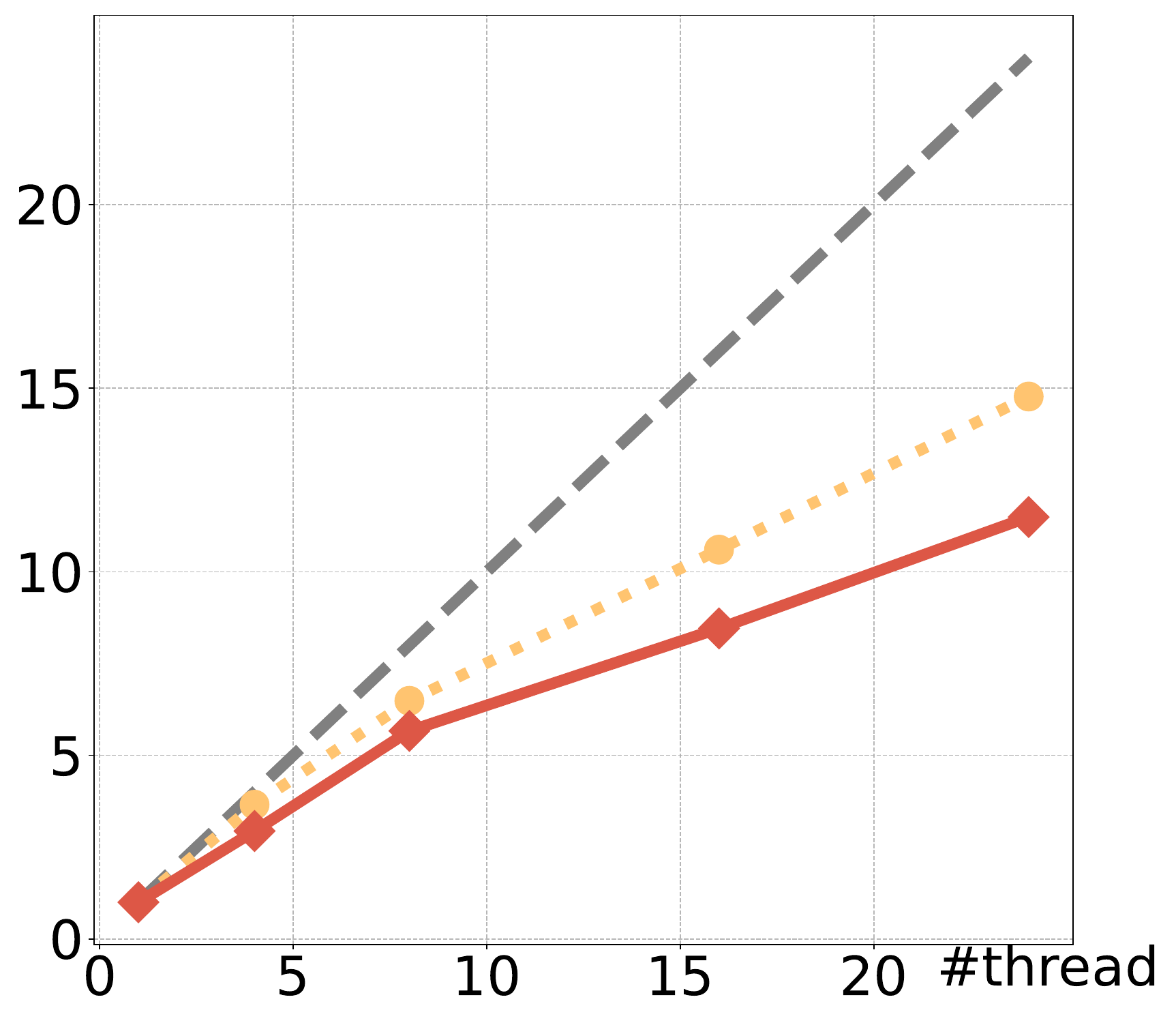}
        \label{fig:lu}
    \end{subfigure}
    \hfill
    \begin{subfigure}{0.48\linewidth}
        \centering
        \captionsetup{skip=-0.2em}
        \caption{IS (local mem=16.10GB)}
        \includegraphics[width=0.9\linewidth]{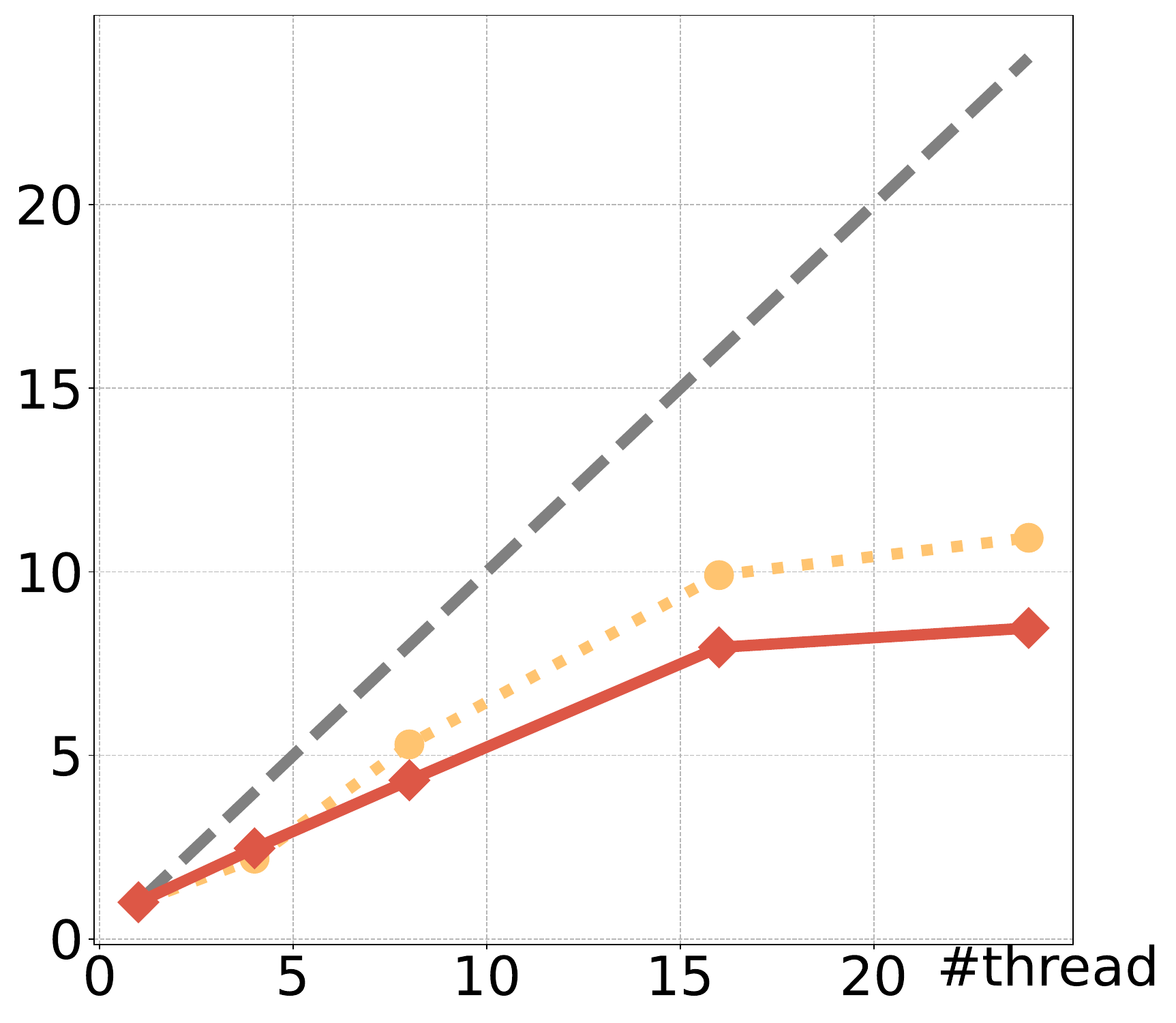}
        \label{fig:is}
    \end{subfigure}
    \vspace{-5pt}
    \caption{Speedup between \texttt{Oracle} and \name with varying thread counts.}
    \vspace{-10pt}
    \label{fig:multi_thread}
\end{figure}

\subsection{Multi-threading}
To evaluate \name's multi-threading performance, we analyze NPB workloads using 1 to 24 threads. For each workload, we set its local memory capacity to the minimum size that achieves comparable performance with a single thread compared to \texttt{Oracle} (where all memory accesses are local), as identified in Section~\ref{sec:overall_performance}. 
We omit XSBench and miniAMR results as these workloads show minimal performance improvement with multiple threads. We focus on more scalable workloads for meaningful insights. 

We evaluate the scalability of both \texttt{Oracle} and \name by plotting their self-normalized speedup curves. In Figure~\ref{fig:multi_thread}, the speedup for \texttt{Oracle} and \name is calculated independently, each normalized to its own single-thread performance. This shows how \texttt{Oracle} and \name scale with increasing thread counts.


The Y-axis in Figure ~\ref{fig:multi_thread} illustrates the scalability of \name across different thread counts for six HPC workloads. We observe that \name generally achieves better or comparable speedup compared to \texttt{Oracle}, particularly as the number of threads increases.

In Figure~\ref{fig:multi_thread}(a,b and d), we observe that CG, MG, and FT show that \name generally achieves better or comparable scalability compared to \texttt{Oracle} as the thread count increases. In particular, \name sustains nearly linear scaling for CG and FT, outperforming \texttt{Oracle} at higher thread counts, while for MG, although both \name and \texttt{Oracle} exhibit sub-linear scaling, Axolotl still maintains a clear advantage at 24 threads. The improved parallel efficiency stems from our two-level RDMA scheduling mechanism, which reduces resource contention as thread count increases by better coordinating remote memory operations across threads.

In contrast, for BT (Figure~\ref{fig:multi_thread}(c)), the scaling of \name significantly lags behind Oracle. While \texttt{Oracle} continues to achieve notable speedup with more threads, \name's performance plateaus early, suggesting that BT’s memory access patterns limit the benefits of remote memory optimization.

For LU and IS (Figure~\ref{fig:multi_thread}(e and f)), both \name and \texttt{Oracle} experience moderate scaling. \name shows a slight advantage over \texttt{Oracle} at higher thread counts for LU, whereas for IS, \name and \texttt{Oracle} exhibit similar trends with \name slightly behind at large core counts. 

Overall, while \texttt{Oracle} benefits from exclusive use of local memory, \name{}’s advanced scheduling and dual buffer design allow it to effectively scale and reduce the performance gap as thread resources increase, particularly in computation-heavy scenarios. Meanwhile, \name does not need to handle synchronization to use multi-threading as our solution is coordinated with OpenMP shared and private data objects.

\subsection{Efficiency of the Dual Buffer Design}
We further analyze the execution time results for different applications to show the efficiency of the dual buffer design in \name. 
For each workload, we set its local memory capacity to the minimum size that achieves comparable performance with a single thread compared to \texttt{Oracle}. 
{Figure~\ref{fig:breakdown} highlights the benefits of the dual buffer approach. Specifically, in CG workloads, the dual buffer optimization significantly reduces execution time compared to configurations without it. These workloads are predominantly read-intensive, making them well suited for the dual-buffer strategy, which overlaps data fetching with computation to effectively hide memory latency. HPC applications often exhibit predictable memory access patterns in large contiguous data blocks, which \name can exploit to maximize the benefits of the dual buffer design.

In contrast, the performance improvements for MG, FT, and LU are less pronounced. These workloads involve a mix of read and write operations that cannot fully utilize the dual buffer design mechanism. The presence of write operations introduces delays that are not effectively mitigated by the dual buffer design, leading to only moderate performance gains.

\begin{figure}[!ht]
    \centering
    \includegraphics[width=0.38
    \textwidth]{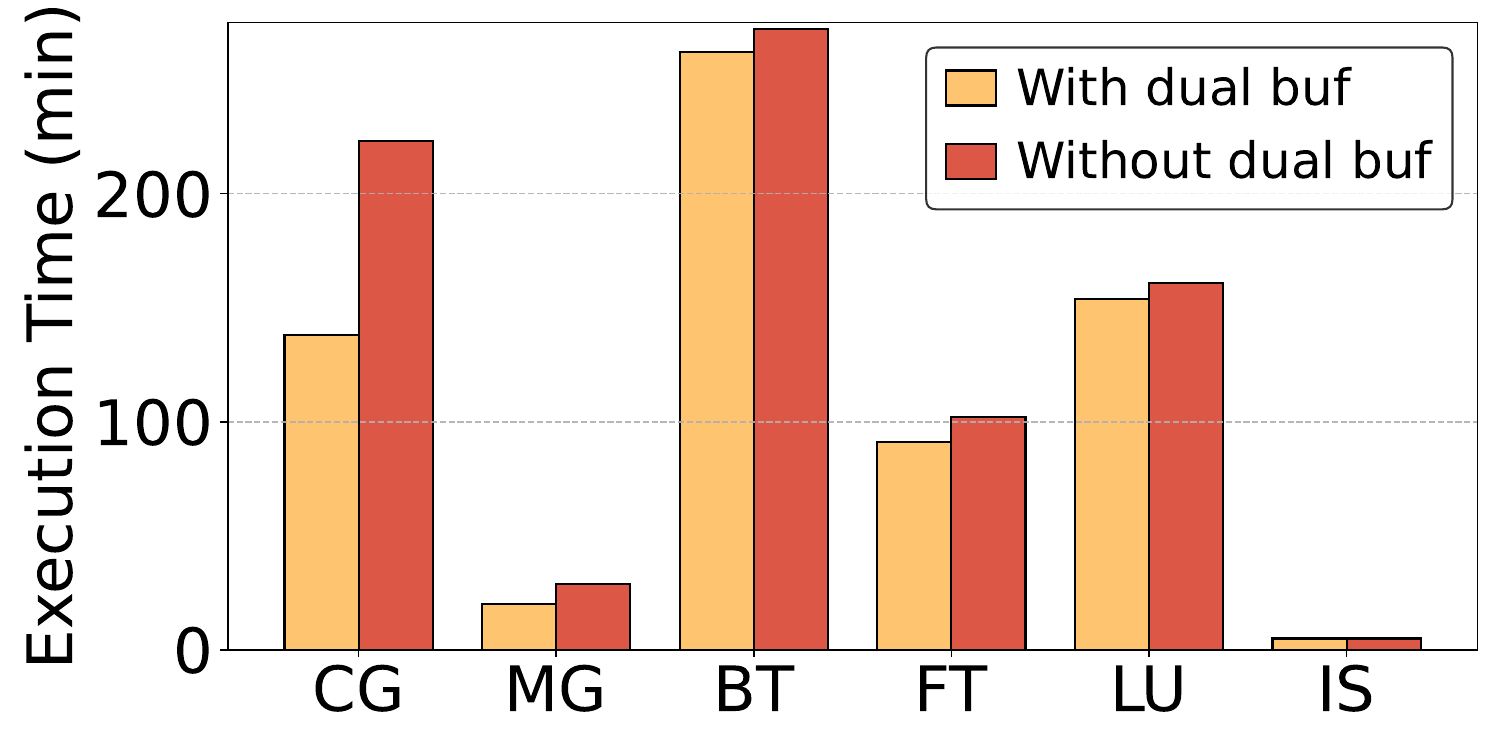}
        \vspace{-5pt}

    \caption{
    Execution time comparison of NPB workloads with and without dual buffer design with a single thread. Lower execution time indicates better performance. 
    }
    \label{fig:breakdown}
\end{figure}

\subsection{Varying Input Problems Sizes}
To assess the performance of \name under different computation intensities, we use CG as an example with varying input problem size with single thread. 
For this test, the local memory size is 0.09 GB. We use this memory size to obtain the results presented in Figure~\ref{fig:sensitivity_size}.
 We compare \name with the \texttt{Oracle} configuration and the scenario that involves synchronous RDMA read and write.

We observe that as the input problem size increases, the throughput gap between \name and the Oracle configuration narrows. For the smallest input sizes (Size S and Size W), \name achieves lower throughput compared to \texttt{Oracle}, mainly due to the overhead associated with RDMA operations, which becomes more significant when the data size is small. This is consistent with the behavior expected, as smaller input sizes do not fully utilize the advantages of RDMA optimizations and often result in higher relative latency overhead.

For large input sizes (Size C and D), \name’s performance shows significant improvement, with throughput levels approaching those of the Oracle configuration. This improvement is attributed to \name’s ability to efficiently overlap computation with data movement, effectively managing larger data exchanges using its dual buffer design.
This confirms that \name is well-suited for handling large-scale HPC workloads, particularly when the input size is substantial enough to benefit from its memory management and data transfer strategies.
This narrowing performance gap with larger input sizes aligns with the findings discussed in Section~\ref{sec:motivation}, where we explored the efficiency of remote memory access for larger data objects. 

\begin{figure}[!ht]
    \centering
    \includegraphics[width=0.35
    \textwidth]{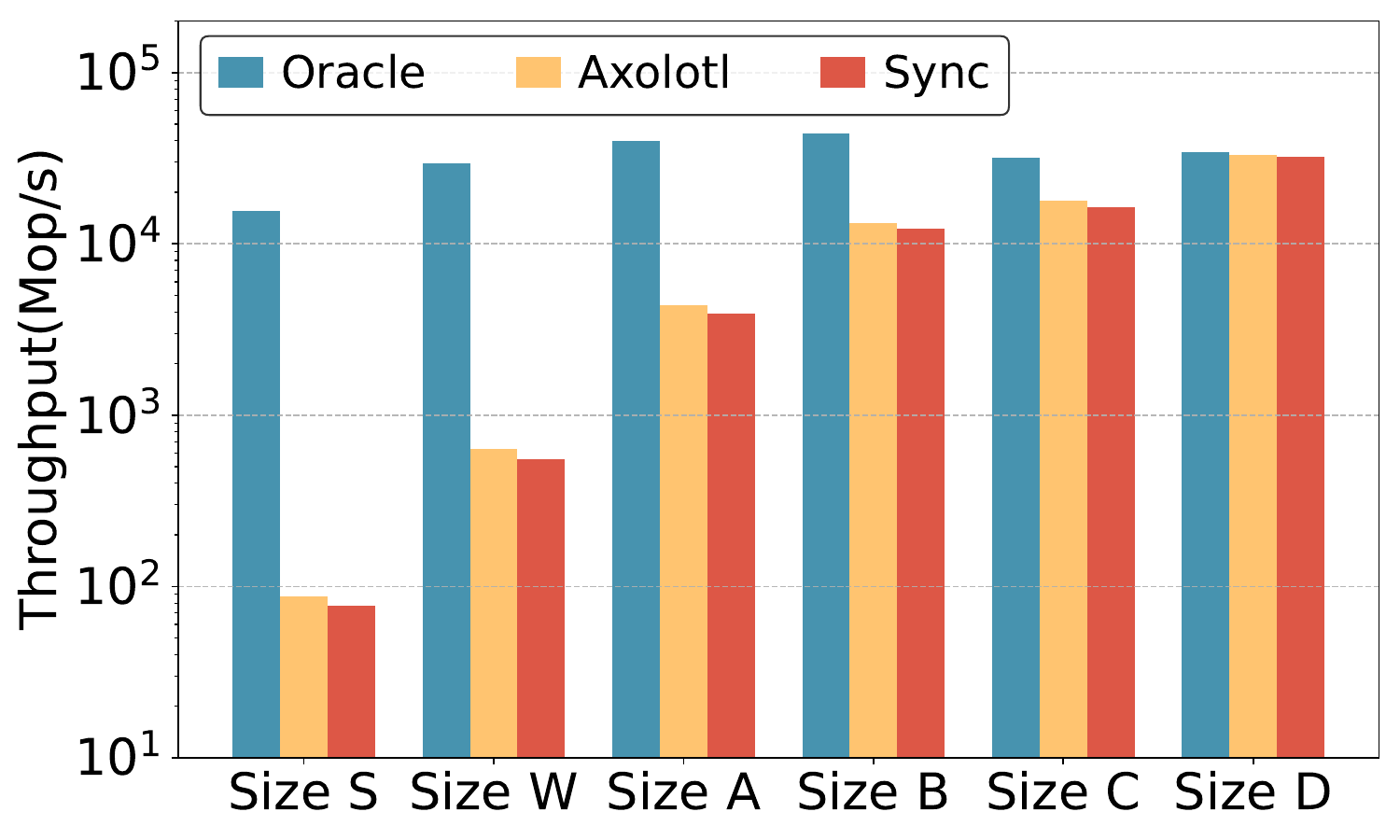}
        \vspace{-5pt}

    \caption{Computation throughput of the CG benchmark with varying input problem sizes.
    }
    \vspace{-10pt}\label{fig:sensitivity_size}
\end{figure}

\section{Related Work}




\subsection{Memory Disaggregation in HPC}
Practitioners explore memory disaggregation to improve memory utilization in HPC centers~\cite{ivy20_hpc_mem_underutilization, ding2023evaluating, sc23_quantitative}. 
Peng et al.~\cite{ivy20_hpc_mem_underutilization} developed a user-space remote-memory paging library to enable applications to explore disaggregated memory on existing HPC clusters. This work examines the potential of throughput scaling with disaggregated memory and sheds light on its applications in the HPC domain.
Ding et al.~\cite{ding2023evaluating} found that most HPC applications either have low memory requirements that can easily fit into a future APU's HBM memory or have a high local-to-remote data movement ratio that prevents local-to-remote bandwidth tapering from impeding performance. This study validates the emerging need for disaggregated memory in HPC applications.  
Wahlgren et al.~\cite{sc23_quantitative} proposed a multilevel quantitative methodology for analyzing application memory system requirements, covering general memory, multi-tier, and memory pooling systems. The proposed work provides a systematic way to understanding and implementing disaggregated memory effectively.
In our work, we address the uniqueness of HPC data and propose a data object-level management to deal with the complex, irregular, and data-dependent access.

%




\subsection{RPC-based Memory Disaggregration} 
Remote Procedure Call (RPC) provides a user-friendly interface to facilitate complex communication operations in large-scale distributed environments. However, the rich features and ease of use come at the cost of high latency and potential responsiveness issues.
FaSST \cite{kalia2016fasst}, a fast RPC framework, is optimized for small data objects but is less suitable for handling large data volumes typical in HPC applications. 
The eRPC \cite{kalia2019datacenter}, a general-purpose RPC library, provides performance comparable to local memory access without relying on RDMA primitives. Despite this, relying solely on RPC often prevents achieving optimal performance in disaggregated memory systems \cite{wang2024rcmp}.
Abrahamse et al. \cite{abrahamse2022memory} introduced an RPC-based object store framework for handling very large datasets. While it achieves competitive memory access speeds compared to local memory and RDMA, performance remains limited by single-thread throughput.
Our proposed framework offers a data object-level memory disaggregation solution that excels in managing large data objects and supports multithreading, making it ideal for HPC applications.

 



\subsection{Offloading Operations to Memory}
Offloading operations to memory reduces the load on computing units by enabling the direct execution of specific tasks (like communication and data processing) within the memory subsystem itself, thereby improving overall data processing efficiency.
Process-In-Memory (PIM) technology~\cite{pawlowski2011hybrid, seshadri2017ambit, kim2021signal, zokaee2018aligner, sun2017energy} embeds processing capabilities directly within memory modules to minimize data movement between computation units (e.g., CPUs) and memory, thereby enhancing processing performance.
Another track to offload operations to memory is by directly transferring tasks to remote memory. For example, Teleport, an offloading system built on LegoOS, a disaggregated operating system architecture, initializes a remote memory region and manages data placement to minimize latency. 
Different from the Compute Express Link (CXL)-based disaggregation that focuses on fast, direct memory access between processors and memory, PIM and Teleport emphasize computing within or near the memory itself.

\section{Conclusion}

We presented \name, a practical data object level memory disaggregation framework designed specifically for HPC applications. \name addresses the challenge of performance overhead in accessing remote memory by identifying and offloading suitable data objects to remote memory, while keeping frequently accessed and performance-critical data objects in local memory. 
Furthermore, the design of \name incorporates a hierarchical memory management strategy, and implementing dual buffer design and asynchronous accesses to hide the latency of remote memory accesses and ensure high performance. 
The evaluation of \name using eight representative HPC workloads and computational kernels demonstrates its effectiveness in achieving practical memory disaggregation. 
The results show that, on average, \name successfully limits the performance degradation to less than 16\% while reducing local memory usage by up to 63\%. 

\bibliographystyle{ACM-Reference-Format}
\bibliography{wdong, ren}

\end{document}